\documentstyle[preprint,pra,aps]{revtex}
\begin{document}
\draft
\title{Renormalization Group Theory of the Three-Dimensional Dilute Bose Gas}
\author{M. Bijlsma and H.T.C. Stoof}
\address{Institute of Theoretical Physics, University of
Utrecht, Princetonplein 5\\ P.O. Box 80.006, 3508 TA Utrecht, The
Netherlands}

\maketitle
\begin{abstract}
We study the three-dimensional atomic Bose gas using renormalization group
techniques. Using our knowledge of the microscopic details of the interatomic
interaction, we determine the correct initial values of our renormalization
group equations and thus obtain also information on nonuniversal properties. As
a result, we can predict for instance the critical temperature of the gas and
the superfluid and condensate density of the Bose-Einstein condensed phase in
the regime $na\Lambda_{th}^2\ll 1$.
\end{abstract}
\vspace{1cm}

\section{Introduction}
After a long history in which a large number of experimental groups around the
world contributed to the development of succesful methods to master
stabilization and cooling of dilute Bose gases, last year the aim of achieving
Bose-Einstein condensation in such a system was finally reached. Indeed, a
macroscopic occupation of the one-particle ground state was irrefutably
observed in magnetically trapped and evaporatively cooled alkali gas samples of
$^{87}$Rb and $^{23}$Na using relatively simple time-of-flight measurements
\cite{rub,nat}. The transition that was claimed to be seen in an experiment
using $^7$Li was less convincing \cite{lit}. In the latter case, the
interatomic interaction is effectively attractive and the potential has a
negative scattering length $a$. Therefore,  Bose-Einstein condensation in this
system is preempted by a first order phase transition to a liquid or solid
phase in the homogeneous case \cite{STONEA}. Nevertheless, for inhomogeneous
gas samples the trapping potential has a stabilizing influence and a
macroscopic occupation of the ground state is possible in principle. However,
when the condensate contains more than some 1500 particles under the conditions
of the $^7$Li experiment \cite{BUDODD}, the condensate is still expected to
collapse \cite{STOTUN}.

After these first experiments, which were primarily aimed at proving the
existence of a Bose condensate, many experimental groups are now building or
improving on their experimental setups to be able to perform much more precise
measurements of various interesting properties of the gas in the degenerate
regime.  Superfluidity \cite{STRING1}, the condensate density and its profile,
the dynamics of condensate formation \cite{STONUCL}, the Josephson effect
\cite{PITA1}, vortex dynamics, collective excitations
\cite{EDWBUR,STRING2,FETTER1,EDWRUP} and the precise value of the critical
temperature are examples of phenomena and quantities of interest. Other types
of experiments will presumably also study the properties of mixtures of atomic
gases. In this respect one might think of two bosonic species with a different
sign of the scattering length, or mixtures of bosons and fermions, or mixtures
uniting both aspects. In the case of a pure fermionic gas of $^6$Li atoms, a
BCS transition to a superfluid state is predicted to occur and should be within
reach of the current experimental technology \cite{STOHOUB}. Furthermore, in
some cases the magnitude and even the sign of the scattering length can be
changed by varying the applied magnetic field. This opens the road to yet
another type of experiment.

It is clear from these possibilities that a large number of experiments are
expected to be performed in the near future, which clearly makes the degenerate
dilute Bose gas also a very interesting subject for theoretical studies.
Indeed, this field of research has rapidly expanded during the last year.
However, most approaches to the dilute Bose gas use the Bogoliubov (or Popov)
theory and are therefore of mean-field type and susceptible to improvements,
both from a practical as well as a fundamental point of view. In these
approaches one mostly uses the so-called two-body $T-$matrix, or equivalently
the scattering length $a$. Technically, this important quantity describes the
collisions taking place in the dilute Bose gas by summing all possible two-body
scattering processes, i.e.\ all ladder diagrams, without taking into account
the fact that the surrounding gaseous medium has an effect on these collisions.
However, we have recently shown that the many-body corrections arising from the
surrounding gas may be important \cite{USVAR}, and are even essential for
solving the problem connected to the order of the phase transition which is
found to be of first order when using the two-body $T-$matrix
\cite{GRIF1,GRIF2,LEEY,STRAL}.

Including quantatively the same many-body corrections in the case of a highly
inhomogeneous gas sample has at this point not yet been done. Moreover,
introducing the effect of the medium on two-particle collisions also in the
condensed phase by means of the many-body $T-$matrix leads to fundamental
problems if we want to describe the physics at long wavelengths correctly as
the ladder diagrams contain infrared divergencies in this case. Using
renormalization group techniques, we expect in principle to be capable of
resolving these infrared problems as with this method a correct resummation of
diagrams is automatically performed, eliminating any potentially troublesome
large distance behavior of the individual diagrams. Furthermore, a
renormalization group calculation can be used to improve the usual mean-field
approaches and the many-body $T-$matrix theory in the critical region. Indeed,
we recently predicted by these means for example that the critical temperature
in the $^{87}$Rb and $^{23}$Na experiments can, due to interaction effects
only, be raised with as much as 10 \% compared to the ideal gas value found
from the criterion $n_c\Lambda_{th}^3=\zeta(3/2)\simeq 2.612$ \cite{USPRL},
which is identical to the criterion also found using mean-field calculations
\cite{USVAR}. Here $n$ is the density and $\Lambda_{th}=(2\pi\hbar^2/
mk_BT)^{1/2}$ the thermal de Broglie wavelength of the atoms in the gas.
Therefore, it is very well conceivable that in the critical region also other
properties of the dilute Bose gas, such as the superfluid and condensate
densities, will significantly change when going beyond the mean-field level or
beyond the many-body $T-$matrix theory.

The renormalization group method is a very powerful method which was in first
instance developed by Wilson \cite{wilson} to study the universal properties of
second order phase transitions. The basic idea is to perform the trace in the
grand canonical partition function $Z_{gr}=Tr(e^{-\beta({\cal H}-\mu{\cal
N})})$ gradually, starting with the high momentum states. After each step one
tries to find a new effective Hamiltonian such that $Z_{gr}=Tr'(e^{-\beta({\cal
H}'-\mu'{\cal N})})$ and the trace is limited to the low momentum states which
have not been reached yet. One proceeds until the complete sum has been
performed. Besides the partition function one in this manner also ends up with
the effective Hamiltonian describing the long distance properties of the
system.

The renormalization group method has been applied to the dilute Bose gas
\cite{FISHOH,WEICHM,KOLSTR1,KOLSTR2} before, but without actually performing an
extensive quantitative study of this system. This is due to the fact that in
general the quantities of nonuniversal nature, such as the critical temperature
and the superfluid and condensate densities, depend on the microscopic details
of the system considered. Put differently, they depend on the ultraviolet
cutoff $\Lambda$ of the theory and this quantity is usually unknown. However,
due to the diluteness of the gas the nonuniversal properties are in the present
and forthcoming experiments the most interesting ones, and therefore we are in
this paper mainly interested in these aspects. The reason that it is
nevertheless possible to perform a quantitative study of the dilute Bose gas
using the renormalization group method, is that for this system we do have
sufficient information about the microscopic details to calculate and predict
the nonuniversal properties. Thus we can, by correctly applying this knowledge,
use the renormalization group method and eliminate the cutoff dependence at the
same time. We will come back to this point in Sec.\ II. Furthermore, we will
show that, in contrast to regular perturbation theory, the problems related to
the infrared divergencies are in principle indeed resolved, but lead
nevertheless to some problems whose solution requires further investigation.
However, these problems are only of importance when the interaction energy is
no longer negligible compared to the kinetic energy of the particles. The
dimensionless parameter reflecting this aspect is $na\Lambda_{th}^2$.
Therefore, we will in this paper first concentrate on the regime where
$na\Lambda_{th}^2\ll 1$.

We treat here only the homogeneous Bose gas with effectively repulsive
interactions, i.e.\ with a positive scattering length. However, as in all
experiments up till now the number of particles $N$ is so large that the
critical temperature $T_c$ is much larger than the energy splitting
$\hbar\omega$ between subsequent levels, one can practically for all
temperatures use a local density approximation to describe the gas in the trap.
The criterion for this description to be valid is that the correlation length
should be smaller than the typical length scale on which the atomic density
varies. Therefore, a local density approximation breaks down close to the edge
of the gas cloud, which is for most practical purposes an unimportant region,
but also in the center of the trap if the temperature approaches the critical
temperature and the diverging correlation length starts to exceed the typical
dimensions of the trap. The temperature interval where this occurs has a width
of $O(T_c(\hbar\omega/ k_BT_c))$ around the critical temperature. As this
region is very small, we conclude that the results we find in this paper for
the homogeneous gas are essentially also valid for the inner part of the
trapped Bose gases, and in particular pertain to the $^{87}$Rb and $^{23}$Na
experiments. Furthermore, we want to remark here that it would in principle
also be possible to set up a renormalization group calculation for the
inhomogeneous Bose gas. Of course, there is no real second order phase
transition present in this system because the correlation length $\xi$ can
never become infinite, but the techniques of renormalization group as presented
in this article can still be used to calculate the (nonuniversal) properties
also in this case.

The paper is organized as follows. In Sec.\ II we briefly discuss the
theoretical description of the dilute Bose gas and the renormalization group
method. In Sec.\ III we first apply the renormalization group to the
uncondensed phase of the Bose gas because the flow equations are relatively
simple and easy to understand in this case. In Sec.\ IV we then go over to the
degenerate Bose gas and again describe the gas properties following from the
renormalization group approach. Finally, in Sec.\ V we end with some concluding
remarks. In the numerical calculations we always use $^{23}$Na as an example,
because the experiment with this atomic species is closest to the conditions of
homogeneity \cite{nat}. We take in these cases the most up-to-date value of $52
a_0$\ for the two-body scattering length \cite{EITETI}.

\section{The Renormalization Group}
The renormalization group equations are most easily obtained using the
functional integral formulation of the grand canonical partition function
\cite{NEGO}. We thus write
\begin{equation}
Z_{gr}=Tr(e^{-\beta({\cal H}-\mu{\cal N})})=\int
d[\psi^*]d[\psi]\exp\left\{-\frac{1}{\hbar}S[\psi^*,\psi]\right\}~.
\end{equation}
This functional integral is over $c$-number fields $\psi^*({\bf x},\tau)$ and
$\psi({\bf x},\tau)$ periodic in imaginary time over $\hbar\beta=\hbar/k_BT$.
The so-called Euclidian action for the dilute Bose gas is given by
\begin{eqnarray}
S[\psi^*,\psi]&=&\int_{0}^{\hbar\beta}d\tau \left( \int d{\bf x}~
\psi^*({\bf x},\tau)\left[ \hbar\frac{\partial}{\partial\tau}-
\frac{\hbar^2{\bf \nabla}^2}{2m}-\mu \right] \psi({\bf x},\tau)\right.
\nonumber \\
&&+\left.\frac{1}{2}\int d{\bf x}\int d{\bf x'}~\psi^*({\bf
x},\tau)\psi^*({\bf
x'},\tau)V({\bf x}-{\bf x'})\psi({\bf x'},\tau)\psi({\bf x},\tau) \right)~,
\end{eqnarray}
with $\mu$ the chemical potential and $V({\bf x}-{\bf x'})$ the
effectively repulsive interaction potential. In principle, the action also
contains a term describing three-particle interactions, i.e.\
\begin{eqnarray}
\frac{1}{6}\int_0^{\hbar\beta}d\tau\int d{\bf x}\int d{\bf x'}\int d{\bf
x''}~\psi^*({\bf x},\tau)\psi^*({\bf x'},\tau)\psi^*({\bf x''},\tau)U({\bf
x}-{\bf x'},{\bf x-x''})\psi({\bf x''},\tau)\psi({\bf x'},\tau)\psi({\bf
x},\tau)~\nonumber,
\end{eqnarray}
and terms describing interactions between four and more particles due to the
finite extent of the electron clouds of the atoms. However, as we are
describing the {\em dilute} Bose gas these terms are expected to make in
general no significant contribution to the thermodynamic properties to be
calculated. Indeed, in usual approaches to the dilute Bose gas these terms are
always neglected for this very reason. One aspect of the renormalization group
calculation is that it is possible, and even rather straightforward, to include
the three-body interaction in the calculations. As its influence will turn out
to be extremely small, except in the critical region where it becomes somewhat
larger, we will in the rest of this section omit this term for reasons of
brevity. However, in Sec.\ III we perform some calculations including the
three-body term to show its effect quantatively, and at that point introduce it
again.

Expanding the fields in Fourier modes through
\begin{equation}
\psi({\bf x},\tau)=\frac{1}{(\hbar\beta V)^{1/2}}\sum_{{\bf k},n} a_{{\bf
k},n}e^{i({\bf k\cdot x}-\omega_n\tau)}
\end{equation}
and the complex conjugate expression for $\psi^*({\bf x},\tau)$, we can write
the action in momentum space as
\begin{eqnarray}
\label{basisactie}
S[a^*,a]=\sum_{{\bf k},n}(-i\hbar\omega_n+\epsilon_{{\bf k}}-\mu)
a^*_{{\bf k},n}a_{{\bf k},n}+\frac{1}{2}\frac{1}{\hbar\beta V}
\sum_{\stackrel{\bf
k,k',q}{n,n',m}} V_{\bf q}a^*_{{\bf k+q},n+m}a^*_{{\bf k'-q},n'-m}
a_{{\bf k'},n'}a_{{\bf k},n}~.
\end{eqnarray}
In this equation $\epsilon_{{\bf k}}=\hbar^2{\bf k}^2/2m$ is the kinetic
energy, $V_{\bf q}=\int d{\bf x}~V({\bf x})e^{-i{\bf q\cdot x}}$ is the Fourier
transform of the interaction potential, $V$ is volume of the system and the
bosonic Matsubara frequencies $\omega_n=2\pi n/\hbar\beta$ reflect the
periodicity of the fields in imaginary time.

The renormalization group equations now follow from repeatedly applying the
renormalization group transformation to this action. It consists of three
different stages \cite{wilson,SHANK} which we summarize here. The first step is
to perform the functional integral in Eq.\ (1) for the most rapidly oscillating
fields. In other words, we eliminate the highest Fourier modes from the action
in Eq.\ (\ref{basisactie}). In practice, we therefore have to split the fields
in `slow modes' and `fast modes', i.e.\ we define
\begin{eqnarray}
\label{modesplitting}
\begin{array}{ll}
\psi_<({\bf x},\tau)=\frac{1}{(\hbar\beta V)^{1/2}}\sum_{{\bf k},n} a_{{\bf
k},n}e^{i({\bf k\cdot x}-\omega_n\tau)} & \mbox{for \ $0<\mid{\bf
k}\mid<\Lambda-d\Lambda$} \\
\psi_>({\bf x},\tau)=\frac{1}{(\hbar\beta V)^{1/2}}\sum_{{\bf k},n} a_{{\bf
k},n}e^{i({\bf k\cdot x}-\omega_n\tau)} & \mbox{for \
$\Lambda-d\Lambda<\mid{\bf k}\mid<\Lambda$}
\end{array}
\end{eqnarray}
where $\Lambda$ is the ultraviolet cutoff in the theory, and d$\Lambda$ is the
thickness of the shell in momentum space which is going to be integrated out.
We will dwell on the issue of the ultraviolet cutoff shortly. With the
definitions in Eq.\ (\ref{modesplitting}), we can write the action as
\begin{eqnarray}
S[\psi^*,\psi]=S_0[\psi^*_<,\psi_<]+S_0[\psi^*_>,\psi_>]+S_I[\psi^*_<,\psi_<,
\psi^*_>,\psi_>] ,
\end{eqnarray}
where $S_0$ is quadratic in the fields, and $S_I$ contains the remaining terms,
and in particular the terms that couple the slow and fast modes. The partition
function can now be written as
\begin{eqnarray}
Z_{gr}&=&\int
d[\psi^*_<]d[\psi_<]\exp\left\{-\frac{1}{\hbar}S_0[\psi^*_<,\psi_<]\right\}
\times\int d[\psi^*_>]d[\psi_>]\exp\left\{-\frac{1}{\hbar}(S_0[\psi^*_>,\psi_>]
+S_I[\psi^*_<,\psi_<,\psi^*_>,\psi_>]\right\}\nonumber \\
&&\equiv\int d[\psi^*_<]d[\psi_<]\exp\left\{-\frac{1}{\hbar}S_0[\psi^*_<,
\psi_<]\right\}\times\exp \left\{-\frac{1}{\hbar}\Delta S[\psi^*_<,\psi_<]
\right\}~.
\end{eqnarray}
After performing the last integral, we obtain an expression for the partition
function containing the new effective action $S'[\psi^*_<,\psi_<]=S_0
[\psi^*_<,\psi_<]+\Delta S[\psi^*_<,\psi_<]$ for the slow modes. The effect of
this mode elimination is to change the value of the existing coupling constants
and to introduce new vertices in the original action. The discussion of the
rules on how we can obtain the equations describing this change of the coupling
constants in general will be deferred to Secs. III and IV where we will also
apply these rules to find the flow equations for the Bose system. The two
remaining stages of the renormalization group transformation consist of a
rescaling of the momenta such that the new cutoff, which is $\Lambda-d\Lambda$,
is restored to its initial value $\Lambda$, and a rescaling of the frequency
and fields in the action such that there is no effect from this momentum
rescaling on some appropriate terms in the quadratic part of the action. If we
neglect the renormalizations from the first step, this procedure yields the
so-called trivial scaling of the coupling constants and reveals the relevance
of the various vertices as it tells whether the value of the corresponding
coupling constant grows (termed relevant), remains the same (termed marginal),
or shrinks (termed irrelevant) as we scale the ultraviolet cutoff. Having
completed these final steps of the renormalization group transformation, we can
read off the new coupling constants from the action. Thus, the renormalization
group equations have two contributions. One is from the rescaling, the other is
from the actual integrating out of high momenta from the action.

Concerning the trivial scaling, there are two cases to distinguish. For the
normal renormalization group equations this scaling is found using the full
quadratic action. The other situation occurs exactly at the critical
temperature. At that point the correlation length $\xi$ of the system is
infinite, and the system looks the same on every length scale. Hence, the
coupling constants remain the same when performing the renormalization group
transformation and we are at a fixed point. To find this fixed point from the
set of renormalization group equations, we generally have to use a different
scaling from the one found above, and this is the scaling at the critical
temperature. It can easily be obtained by realizing that the correlation time
$\tau_c$ also diverges as we approach the critical temperature, behaving like
$\tau_c\propto \xi^z$. At temperatures such that $\hbar\beta\ll \tau_c$, we can
neglect the time dependence in the action and the $\partial/\partial\tau -$term
in the quadratic part of the action is unimportant. As a result, we then only
need to consider the kinetic energy term to find the trivial scalings. This
yields the scaling at the critical temperature. In renormalization group
studies, one normally restricts oneself to couplings which are relevant or
marginal at the critical temperature, and this we will in first instance also
do. However, it turns out that even the marginal coupling constant $U_{\bf 0}$
from the three-body interaction term is quite irrelevant to the nonuniversal
properties to be calculated. This is of course due to the fact that three-body
interactions are unimportant for a {\em dilute} system. However, if one is
interested in the universal properties of the phase transition, such as e.g.\
the critical exponent $\nu$ connected to the divergence of the correlation
length when approaching the critical temperature, marginal and also irrelevant
couplings can have a considerable effect. We will explicitly encounter this
fact in the next section. Also the position of the fixed point is changed when
including irrelevant couplings.

The renormalization group equations describe the change of the coupling
constants as we integrate out momentum shells from the action. Having arrived
at the long wavelength effective action, we find whether or not we are in the
condensed phase, and we can calculate universal properties connected to this
phase transition. However, our aim is to use the renormalization group method
to obtain information on other properties as well. In particular, we want to
calculate the equation of state, the superfluid density, and the pressure of
the gas. This can also straightforwardly be done by noting that e.g.\ for the
total density we have $n=\sum_{{\bf k},n}\langle a^*_{{\bf k},n}a_{{\bf
k},n}\rangle$. This equation can be cast into a differential equation
describing the building up of the density as we proceed with the elimination of
the fast Fourier modes. For that purpose, we have to use the right value of the
chemical potential $\mu$ in each subsequent momentum shell as found from the
renormalization group equations. In the same fashion one can also determine
e.g.\ the superfluid density and the thermodynamic potential.

Before we can start with the derivation of the flow equations, we first have to
pay some attention to the high momentum limit of the action. In principle,
there is no real sharp ultraviolet cutoff $\Lambda$ in our problem. However,
the typical behavior of the Fourier transform of the two-body interaction
potential, depicted in Fig.\ 1, is such that there is an effective ultraviolet
cutoff around the momentum scale set by the scattering length $a$ of this
potential. Below this value, the Fourier transform is practically momentum
independent and equal to $V_{\bf 0}$. As in the Bose systems considered here
and realized experimentally the temperatures are so low that
$\hbar/a\gg\hbar/\Lambda_{th}$, the particles in the gas reside in a momentum
range well below this ultraviolet cutoff. Thus, we can represent the
interaction potential by the momentum independent value $V_{\bf 0}$ for momenta
below a cutoff $\hbar\Lambda$ of $O(\hbar/a)$, and zero for larger momenta.

Modelling the potential as such implies that the nonuniversal properties we
find from a renormalization group calculation will be sensitive to the specific
value of the cutoff $\Lambda$ taken in the calculations. However, at this point
our knowledge about the microscopic details of the Bose gas comes in to resolve
this potential problem. In particular, we know that the two-body interaction
potential $V_{\bf q}$ has to renormalize to the two-body $T-$matrix
$T^{2B}(({\bf k-k'})/2+{\bf q},({\bf k-k'})/2;\hbar^2({\bf k-k'})^2/m)$ when we
include all possible two-body scattering processes in the vacuum \cite{GLOC}.
The two-body $T-$matrix roughly has the same momentum dependence as $V_{\bf
q}$, cf.\ Fig.\ 1, and is in particular constant and equal to $4\pi a\hbar^2/m$
in the range of thermal momenta and energies. Thus, given an ultraviolet cutoff
$\Lambda$ we can fix the renormalization group equations by demanding that for
the two-body problem, $V_{\bf 0}$ indeed correctly renormalizes to $4\pi
a\hbar^2/m$. Since this value is, due to the inequality $a/\Lambda_{th}\ll 1$,
already attained before we enter the thermal regime as we integrate out more
and more momentum shells from the action, this indeed leads to a correct
description of the properties of the Bose gas which is independent of the
ultraviolet cutoff $\Lambda$. Having eliminated the cutoff dependence, we are
then in a position to determine also the nonuniversal properties of the dilute
Bose gas. Furthermore, we can perform the calculation for any (positive) value
of the scattering length, thus being able to describe any atomic species with
effectively repulsive $s-$wave scattering. The results we find are therefore
relevant to the current experiments using $^{87}$Rb and $^{23}$Na, but also to
future experiments using atomic hydrogen or other atoms with a positive
scattering length.

We now turn to the application of the renormalization group method to the
dilute Bose gas. First, we derive the renormalization group equations when the
chemical potential is negative, starting from the action in Eq.\
(\ref{basisactie}). These equations however, do not describe the Bose condensed
phase for then it is required that the chemical potential be positive. As a
result, we have to rederive the renormalization group equations for that case
and this is carried out in Sec.\ IV. The derivation is now much more involved
as the space and time independent part of the effective action, i.e.\ $-\mu\mid
a_{{\bf 0},0}\mid^2+V_{\bf 0}\mid a_{{\bf 0},0}\mid^4/2\hbar\beta V$, has for
$\mu>0$ a Mexican hat shape and we first have to expand the action around the
correct extremum by performing the shift $a_{{\bf 0},0}\rightarrow a_{{\bf
0},0}+\sqrt{n_{\bf 0}\hbar\beta V}$ and introducing the condensate density
$n_{\bf 0}$. Only after that can we proceed to find the contributions to the
renormalization of the vertices.

\section{The symmetry unbroken phase}
In this section we first concentrate on the renormalization group equations
valid for negative chemical potential. The reasons for this are threefold. The
equations correctly describe the Bose gas in the uncondensed phase, and they
offer an easy way to determine the influence of three-body effects on the
quantities of interest. Moreover, it is best to start with this relatively
simple set of equations because it illuminates most clearly our procedure to
eliminate the cutoff dependence of the theory.

\subsection{The flow equations}
To calculate the change of the couplings after each step of the renormalization
group transformation we can technically proceed in two different, but
equivalent ways. The first and probably most familiar method is to expand the
integrand in the partition function in powers of
$S_I[\psi^*_<,\psi_<,\psi^*_>,\psi_>]$, and then perform the integrals over the
fast modes by evaluating the appropriate Feynman diagrams contributing to the
renormalization or flow of the vertices of interest. The renormalization of the
gradient and time-derivative terms of interest are found by performing a Taylor
expansion in external momenta and frequencies of these diagrams. However, the
associated couplings become more and more irrelevant for higher order terms in
this expansion. The integration over the internal momenta of the diagrams is
restricted to the afore-mentioned momentum shell, which can in principle be
chosen to have any thickness d$\Lambda$. Taking it infinitesimally small,
possible in the thermodynamic limit only, leads to a continuous renormalization
group transformation described by a coupled set of first order differential
equations. A much more important reason to take d$\Lambda$ infinitesimal is
that in this case the type of diagrams which contribute to the renormalization
are the so-called one-loop diagrams only. This is a point which is not always
explicit in the literature, but it follows from the fact that each extra loop,
i.e.\ each extra momentum integral, introduces an extra factor of d$\Lambda$
\cite{HUANG}, and therefore will vanish when d$\Lambda$ is infinitesimal.
Clearly, this is not so when the thickness d$\Lambda$ has a finite value. Note
furthermore that the above holds
irrespective of the magnitude of the interactions, in contrast to what is
sometimes mentioned in the literature \cite{FISHOH,KOLSTR1}. In the case of
e.g.\ a large two-body interaction vertex, the infinitesimal thickness of the
momentum shell still ascertaines the validity of a perturbation expansion with
the use of one-loop diagrams only. This most notably implies that in this
formulation the restrictions of the renormalization group method lie in the
number of vertices considered, and not in the type of diagrams taken into
account. The second way to obtain the renormalization group equations does not
explicitly make use of Feynman diagrams, and is therefore very useful and
efficient when the number of diagrams is large and/or the associated
combinatorics is complicated. We here use the diagrammatic method as in the
unbroken phase the number of diagrams is limited and the combinatorics is
simple. We use the other method in Sec.\ IV. Moreover, using Feynman graphs
also gives a transparent way to find the trivial scaling of the vertices in a
somewhat different fashion than from the rescaling procedure described in Sec.\
II, which brings out more clearly the physics of this procedure.

As mentioned before, we include in first instance also the three-body
interaction term written down in the previous section into our considerations,
and take it, as the two-body interaction, to be momentum independent below the
cutoff $\Lambda$. The one-loop diagrams to be calculated are depicted in Fig.\
2. We find that in the thermodynamic limit the total contribution to the
chemical potential from integrating out an infinitesimal momentum shell in the
Hartree and Fock diagrams is
\begin{eqnarray}
\label{mu1loop}
d\mu=-2V_{\bf 0}\int^{\Lambda(l)}_{\Lambda(l)-d\Lambda}\frac{d^3{\bf
k}}{(2\pi)^3}N(\epsilon_{\bf k}-\mu)~,
\end{eqnarray}
where $\Lambda(l)=\Lambda e^{-l}$ denotes the radius of the shell in momentum
space. The ladder and bubble diagrams renormalizing the two-body interaction
potential give, together with the diagram containing the three-body term,
\begin{eqnarray}
dV_{\bf 0}=-V_{\bf 0}^2\int^{\Lambda(l)}_{\Lambda(l)-d\Lambda}\frac{d^3{\bf
k}}{(2\pi)^3}\frac{1+ 2N(\epsilon_{\bf k}-\mu)}{2(\epsilon_{\bf
k}-\mu)}-4V_{\bf 0}^2\int^{\Lambda(l)}_{\Lambda(l)-d\Lambda} \frac{d^3{\bf
k}}{(2\pi)^3}\beta N(\epsilon_{\bf k}-\mu)[N(\epsilon_{\bf k}-\mu)+1]+
\nonumber \end{eqnarray}
\begin{eqnarray}
\label{vnul1loop}
3U_{\bf 0}\int^{\Lambda(l)}_{\Lambda(l)-d\Lambda}\frac{d^3{\bf k}}{(2\pi)^3}
N(\epsilon_{\bf k}-\mu)~,
\end{eqnarray}
and for the vertex $U_{\bf 0}$ we find
\begin{eqnarray}
dU_{\bf 0}=8V_{\bf 0}^3\int^{\Lambda(l)}_{\Lambda(l)-d\Lambda}\frac{d^3{\bf
k}}{(2\pi)^3}\beta^2 N(\epsilon_{\bf k}-\mu)[N(\epsilon_{\bf
k}-\mu)+1][2N(\epsilon_{\bf k}-\mu)+1]+ \nonumber
\end{eqnarray}
\begin{eqnarray}
3V_{\bf 0}^3\int^{\Lambda(l)}_{\Lambda(l)-d\Lambda}\frac{d^3{\bf
k}}{(2\pi)^3}\frac{1}{(\epsilon_{\bf k}-\mu)^2}[1+2N(\epsilon_{\bf
k}-\mu)+2\beta (\epsilon_{\bf k}-\mu) N(\epsilon_{\bf k}-\mu)[N(\epsilon_{\bf
k}-\mu)+1]]- \nonumber
\end{eqnarray}
\begin{eqnarray}
\label{unul1loop}
U_{\bf 0}V_{\bf 0}\int^{\Lambda(l)}_{\Lambda(l)-d\Lambda}\frac{d^3{\bf
k}}{(2\pi)^3}[\frac{3(1+2N(\epsilon_{\bf k}-\mu))}{\epsilon_{\bf
k}-\mu}+18\beta N(\epsilon_{\bf k}-\mu)[N(\epsilon_{\bf k}-\mu)+1]~.
\end{eqnarray}
In these expressions $N(\epsilon_{\bf k}-\mu)=1/(e^{\beta(\epsilon_{\bf
k}-\mu)}-1)$ is the Bose-Einstein distribution function which result from the
summation over the Matsubara frequencies $\omega_n$.

To derive the renormalization group equations and the trivial scalings of the
coupling constants we focus on the first of these equations renormalizing the
chemical potential, and cast it into a differential equation. Using $\mid {\bf
k}\mid=\Lambda(l)=\Lambda e^{-l}$ and performing the angular integrals, we can
write Eq.\ (\ref{mu1loop}) as
\begin{eqnarray}
\label{v0bovenstep1}
d\mu=-2V_{\bf 0}\frac{\Lambda^3}{2\pi^2}\int^{l+dl}_{l} N(\epsilon_{\Lambda}
e^{-2l}-\mu)e^{-3l}dl~.
\end{eqnarray}
Next, we remove all explicit $l-$dependencies from the Bose-Einstein
distribution function by letting both the temperature and the chemical
potential scale with exponent 2, i.e.\ we put $T(l)=Te^{2l}$ and $\mu (l)=\mu
e^{2l}$. Hence, both temperature and chemical potential scale trivially with
exponent 2 and we find for the differential equation describing the change of
the chemical potential when integrating out a momentum shell
\begin{eqnarray}
\frac{d\mu}{dl}=2\mu-\frac{\Lambda^3}{\pi^2}V_{\bf 0}e^{-l}N(\epsilon_{\Lambda}
-\mu)~.
\end{eqnarray}
Finally, we now also absorb the factor $e^{-l}$ into $V_{\bf 0}$ in order to
remove the remaining explicit $l-$dependence. As a result, $V_{\bf 0}$ scales
trivially with exponent $-1$. By considering in Eq.\ (\ref{vnul1loop}) the term
containing $U_{\bf 0}$, we can analogously show that this vertex trivially
scales with an exponent $-4$. Note that to find the real physical quantities we
should always remove the trivial scalings again. We see that the trivial
scaling can, in a very simple way, be found from the one-loop expressions.
Moreover, it shows that introducing the trivial scalings does not have an
essential effect on the renormalization of the vertices; it is merely a
rewriting of the differential equations. In the case of a negative chemical
potential we eventually obtain the following coupled set of renormalization
group equations for the coupling constants $\mu, V_{\bf 0}$ and $U_{\bf 0}$,
\begin{mathletters}
\begin{eqnarray}
\label{murgboven}
\frac{d\mu}{dl}&=&2\mu-\frac{\Lambda^3}{\pi^2}V_{\bf 0}N(\epsilon_{\Lambda}
-\mu)~,
\end{eqnarray}
\begin{eqnarray}
\label{vnulrgboven}
\frac{dV_{\bf 0}}{dl}&=&-V_{\bf 0}-\frac{\Lambda^3}{2\pi^2}V_{\bf 0}^2\left[
\frac{1+2N(\epsilon_{\Lambda}-\mu)}{2(\epsilon_{\Lambda}-\mu)}+4\beta
N(\epsilon_{\Lambda}-\mu)[N(\epsilon_{\Lambda}-\mu)+1]\right]\nonumber \\
&&+\frac{3\Lambda^3}{2\pi^2}U_{\bf 0}N(\epsilon_\Lambda-\mu)~,
\end{eqnarray}
\begin{eqnarray}
\label{unulrgboven}
\frac{dU_{\bf 0}}{dl}&=&-4U_{\bf 0}+\frac{\Lambda^3}{2\pi^2}V_{\bf 0}^3
\left[8\beta^2N(\epsilon_\Lambda-\mu)[N(\epsilon_\Lambda-\mu)+1][2N(\epsilon_
\Lambda-\mu)+1]+\right. \nonumber \\
&&\left.\frac{3}{(\epsilon_\Lambda-\mu)^2}[1+2N(\epsilon_\Lambda-\mu)+2\beta
(\epsilon_\Lambda-\mu)N(\epsilon_\Lambda-\mu)[N(\epsilon_\Lambda-\mu)+1]]
\right]- \nonumber \\
&&\frac{\Lambda^3}{2\pi^2}U_{\bf 0}V_{\bf 0}\left[\frac{3(1+2N(\epsilon_\Lambda
-\mu))}{\epsilon_\Lambda-\mu}+18\beta N(\epsilon_\Lambda-\mu)[N(\epsilon_
\Lambda-\mu)+1]\right]~.
\end{eqnarray}
\end{mathletters}

To argue that these equations are the only ones we need to consider, we still
have to determine the trivial scaling at the critical temperature. This
different scaling comes about because in the limit $l\rightarrow\infty$ the
Bose-Einstein distribution function behaves as $N(\epsilon_\Lambda-\mu)=1/
\beta(l)(\epsilon_\Lambda-\mu(l))$ as we are effectively at very high
temperatures due to $\beta(l)=\beta e^{-2l}$. To remove again all explicit
$l-$dependencies after substituting this behavior, we clearly need a different
trivial scaling of the vertices, and this is precisely the trivial scaling at
the critical temperature, since putting $N(\epsilon_\Lambda-\mu)=1/\beta
(\epsilon_\Lambda-\mu)$ is equivalent to neglecting the time-derivative in the
action. In this manner we can straightforwardly show that the scaling of the
chemical potential remains the same, i.e.\ we have $\mu(l)=\mu e^{2l}$, that
the scaling of the two-body interaction becomes $V_{\bf 0}(l)=V_{\bf 0}e^{l}$
instead of $V_{\bf 0}(l)=V_{\bf 0}e^{-l}$, and that the three-body interaction
does not scale, i.e.\ $U_{\bf 0}(l)=U_{\bf 0}$ instead of $U_{\bf 0}(l)=U_{\bf
0}e^{-4l}$. From this we conclude that $\mu$ and $V_{\bf 0}$ are relevant and
$U_{\bf 0}$ is marginal at the critical temperature. Furthermore, the four-body
interaction is indeed irrelevant and therefore not included in the
calculations. The coefficients of the gradient and time derivative terms in the
quadratic part of the action are, like $U_{\bf 0}$, marginal and would in
principle also have to be included in the renormalization group equations.
However, as we are in the regime $a/\Lambda_{th}\ll 1$, the interactions are
expected to be independent of momentum and energy (but see below) and there is
no renormalization of the $\partial/\partial\tau$ and $\mid\nabla\mid^2$ terms.
Moreover, we know from the $\epsilon-$expansion that the anomalous dimension
$\eta$, indicating the importance of the $\mid\nabla\mid^2$ renormalization at
the critical temperature, is very small, namely $\eta=0.02$ \cite{ZINNJ}.
Therefore, the $\partial/\partial\tau$ and $\mid\nabla\mid^2$ renormalizations
will be neglected and Eq.\ (13) describes the renormalization of the vertices
we will consider.

To calculate the partition function of the gas, we need, next to the flow
equations for the above quantities, also the correct boundary conditions. The
first one for the chemical potential $\mu$ is just the bare value in the action
Eq.\ (\ref{basisactie}). For the two-body interaction potential $V_{\bf 0}$ we
need to be more careful, as this vertex has to correctly fix the
renormalization group equations as described in the previous section. From Eq.\
(\ref{vnulrgboven}) we recognize that in a vacuum, i.e.\
$N(\epsilon_\Lambda-\mu)=0$, the renormalization of the interaction between two
particles is governed by
\begin{eqnarray}
\label{lipschwing}
\frac{dV_{\bf 0}}{dl}=-V_{\bf 0}-\frac{\Lambda^3}{2\pi^2}V_{\bf
0}^2\frac{1}{2(\epsilon_{\Lambda}-\mu)}~.
\end{eqnarray}
This is just the differential form of the Lippmann-Schwinger equation
\cite{GLOC} for the two-body $T-$matrix at energy $2\mu$, i.e.\ $T^{2B}({\bf
0},{\bf 0};2\mu)$. As the two-body $T-$matrix is energy independent for low
energies, we can neglect $\mu$ and use $T^{2B}({\bf 0},{\bf 0};2\mu)\simeq
T^{2B}({\bf 0,0};0)=4\pi a\hbar^2/m$. As the solution of Eq.\
(\ref{lipschwing}) is also practically independent of the chemical potential,
we can there also neglect it. As a result, the requirement is now that, given
an ultraviolet cutoff $\Lambda$, $V_{\bf 0}$ flows for $l\rightarrow\infty$ to
the value $4\pi a\hbar^2/m$. This can be ascertained by choosing the right
initial condition for $V_{\bf 0}$, and more precisely we find from analytically
solving Eq.\ (\ref{lipschwing}) that
\begin{eqnarray}
V_{\bf 0}(l=0)=\frac{4\pi a\hbar^2}{m}\frac{1}{1-2a\Lambda/\pi}
\end{eqnarray}
leads to the correct result. Note that we can describe different atomic species
by only changing the value of the scattering length $a$ used in this equation.
Finally, we in principle also need a boundary condition for the three-body
interaction $U_{\bf 0}$. For this interaction we can, analogous to the
scattering length $a$ for the two-body interaction, introduce a length scale
$b$, and again fix $U_{\bf 0}(l=0)$ such that the renormalization group
calculation gives the correct result $U_{\bf 0}(l=\infty)=4\pi\hbar^2 b^4/m$
for elastic three-body scattering in a vacuum. However, in general not much is
known about the microscopic details of the three-body interaction in a dilute
Bose gas, and in particular about the value of the `three-body scattering
length' $b$. But, as can be expected from the fact that the three-body
interactions are in the renormalization group sense irrelevant at large
momenta, the results are practically insensitive to the boundary value of
$U_{\bf 0}$ used, and $U_{\bf 0}(l=0)$ is hardly of any importance. This is
shown explicitly in Sec.\ IIIB where we analyze the results from the
renormalization group approach for $^{23}$Na. Note, that taking $U_{\bf
0}(l=0)=0$ is equivalent to assuming that three-particle scattering is solely
due to the sum of pair interactions. This is a standard approximation in atomic
three-body calculations.

Finally, to describe the dilute Bose gas we still need to derive expressions
for the total and superfluid density, and the equation giving the thermodynamic
potential $\Omega$, and thus the pressure $p=-\Omega/V\equiv -\omega$. To do
so, we make use of the following well-known one-loop expressions for the
density
\begin{eqnarray}
n=\int^{\Lambda}_0\frac{d^3{\bf k}}{(2\pi)^3}N(\epsilon_{\bf k}-\mu)~,
\end{eqnarray}
the superfluid density
\begin{eqnarray}
n_s=n-n_n=n-\int^{\Lambda}_0\frac{d^3{\bf k}}{(2\pi)^3}\frac{2}{3}
\beta\epsilon_{\bf k}N(\epsilon_{\bf k}-\mu)[N(\epsilon_{\bf k}-\mu)+1]~,
\end{eqnarray}
where $n_n$ is the normal density given by the momentum-momentum correlation
function, and the thermodynamic potential
\begin{eqnarray}
\omega=\frac{1}{\beta}\int^{\Lambda}_0\frac{d^3{\bf k}}{(2\pi)^3}ln(1-e^{-\beta
(\epsilon_{\bf k}-\mu)})~.
\end{eqnarray}
Casting these equations into a differential form by performing the integration
shell by shell leads to
\begin{mathletters}
\begin{eqnarray}
\frac{dn}{dl}&=&\frac{\Lambda^3}{2\pi^2}N(\epsilon_{\Lambda}-\mu)e^{-3l}
\end{eqnarray}
\begin{eqnarray}
\frac{dn_s}{dl}&=&\frac{dn}{dl}-\frac{\Lambda^3}{3\pi^2}\beta\epsilon_{\Lambda}N(\epsilon_\Lambda-\mu)[N(\epsilon_\Lambda-\mu)+1]e^{-3l}
\end{eqnarray}
\begin{eqnarray}
\frac{d\omega}{dl}&=&\frac{1}{\beta}\frac{\Lambda^3}{2\pi^2}ln(1-e^{-\beta
(\epsilon_\Lambda-\mu)}) e^{-5l}~,
\end{eqnarray}
\end{mathletters}
where the inverse temperature again scales as $\beta(l)=\beta e^{-2l}$ and the
chemical potential is found from Eq.\ (\ref{murgboven}) at each step of the
integration. These equations describe the building up of these quantities as we
integrate out more and more momentum shells from the action. For convenience,
the explicit $l-$dependence is not removed from these equations, and they thus
immediately yield the physical quantities. Note furthermore that these
equations have no influence on the renormalization of $\mu, V_{\bf 0}$, and
$U_{\bf 0}$ as they are decoupled from the renormalization group equations
(13).

\subsection{Analysis of the flow equations}
We start our analysis by first focussing on the critical properties of Eq.\
(13) and in particular on the critical exponent $\nu$ pertaining to the
divergence of the correlation length, i.e.\ the correlation length behaves as
$\xi=\xi_0 \mid(T-T_c)/T_c\mid^{-\nu}$ when approaching the critical
temperature. For that purpose we have to find the fixed point of the
renormalization group equations, linearize the flow equations around this fixed
point and identify the largest eigenvalue $\lambda_+$ which is related to this
critical exponent via $\nu=1/\lambda_+$ \cite{wilson}. We in first instance
omit the three-body interaction, but in a subsequent calculation include it
again to determine the influence of this marginal vertex on the value of $\nu$.
Only for the set \{$\mu,V_{\bf 0}$\}, when $U_{\bf 0}(l)=0$, do we perform the
calculation of the fixed point explicitly. With the remarks made in Sec. IIIA
we have that the fixed point is found from
\begin{mathletters}
\begin{eqnarray}
\label{fipomu}
\frac{d\mu}{dl}=2\mu-\frac{\Lambda^3}{\pi^2}V_{\bf 0}\frac{k_BT}{\epsilon_
\Lambda-\mu}=0
\end{eqnarray}
\begin{eqnarray}
\label{fipovnul}
\frac{dV_{\bf 0}}{dl}=V_{\bf 0}-\frac{\Lambda^3}{2\pi^2}V_{\bf 0}^2
\frac{5k_BT}{(\epsilon_\Lambda-\mu)^2}=0
\end{eqnarray}
\end{mathletters}
yielding $(\mu^*,V_{\bf
0}^*)=(\epsilon_\Lambda/6,5\pi^2\epsilon^2_\Lambda/18k_B T\Lambda^3)$. From
linearization of the Eqs.\ (\ref{fipomu}) and (\ref{fipovnul}) around this
value we find for the largest eigenvalue $\lambda_+=1.878$, implying that
$\nu=0.532$. Repeating the calculation including the equation for $U_{\bf
0}(l)$, the fixed point is shifted and the critical exponent is found to be
$\nu=0.613$. Thus, we see that the marginal operator $U_{\bf 0}$ has a rather
large effect. Moreover, we can conclude from this result that also irrelevant
coupling constants must have a considerable effect as it is known, from the
$\epsilon-$expansion of the $O(2)-$model \cite{ZINNJ} and from measurements in
$^4$He experiments, that the true critical exponent of the Bose gas should have
the value $\nu=0.67$. This discrepancy should be alleviated by including more
and more irrelevant vertices.

However, as we are in particular interested in the nonuniversal properties of
the dilute Bose gas, we now turn to the influence of the three-body interaction
term $U_{\bf 0}$ on these quantities. The influence of this term will of course
be largest when we start close to the critical chemical potential, because at
the critical point in principle all fluctuations are of importance. Starting
with a chemical potential near the critical value leads to a trajectory that
almost flows into the fixed point of the renormalization group equations, and
the momentum interval in which irrelevant vertices can have a significant
contribution to the flow, and also to the building up of the density, is then
largest. The bare chemical potential yielding a flow into the fixed point is
positive, and the physical chemical potential (i.e.\ with the trivial scaling
removed) renormalizes to zero. For a bare chemical potential larger than this
critical value the flow is no longer defined as at some value of $l$ we have
that $(\epsilon_\Lambda-\mu(l))$ becomes zero and the Bose-Einstein
distribution function diverges. We will come back to this point later on and
restrict ourselves here to the accessible regime, which physically implies that
$n<n_c$.

The first aspect connected to the three-body interaction concerns the initial
value problem for $U_{\bf 0}$. Indeed, as alluded to before, an explicit
calculation shows that changing the boundary condition for $U_{\bf 0}$ from 0
to one corresponding with a `three-body scattering length' $b=10a=520a_0$,
which is extremely large in general, changes the total density and the pressure
in the system with less than 0.1 \%.  Thus, the results we obtain are
practically insensitive to this boundary condition and henceforth we simply use
$b=a=52a_0$. The second aspect we want to consider is the influence of the
three-body interaction term itself on the outcome of the renormalization group
flow. This we do by alternatively including and excluding this vertex. That is,
we solve the set $\{\mu,V_{\bf 0},U_{\bf 0}\}$ and the set $\{\mu,V_{\bf 0}\}$
and compare the results we find. For that purpose, we plot in Fig.\ 3 the
$p-n^{-1}-$diagram near the critical density $n_c$, where the influence of
$U_{\bf 0}$ is largest. From this figure we see that the change in density and
pressure is about 1\% at maximum. Far away from the critical conditions, i.e.\
at large negative chemical potential, the system becomes more and more dilute,
and the influence of the $U_{\bf 0}$ term vanishes, consistent with
expectations. In principle, we could choose to maintain the three-body
interaction term in the renormalization group equations. However, as its effect
is very small we will from now on neglect $U_{\bf 0}$ altogether. Thus, the
dilute Bose gas, and more in particular the nonuniversal properties we are
interested in, is accurately described by only following the renormalization of
the chemical potential and the two-body interaction. Having concluded this, we
restrict ourselves from now on to the coupled Eqs.\ (\ref{murgboven}) and
(\ref{vnulrgboven}).

However, before we analyze some physical implications of these equations, we
want to remark that the dependence on the ultraviolet cutoff $\Lambda$ is
indeed eliminated from the theory. The influence on e.g.\ the density can be
shown to be completely absent, of course with the limitations that
$\hbar\Lambda$ should be larger than the thermal momentum $\hbar/\Lambda_{th}$
and that $V_{\bf 0}$ is properly renormalized to $4\pi a\hbar^2/m$ when we
enter the thermal regime. We are going to compare the results from the
renormalization group calculation with known results for the weakly-interacting
Bose gas as found from the many-body $T-$matrix theory \cite{USVAR,USNIST}. Far
from the critical temperature we expect the results of the mean-field and
renormalization group calculations to be identical. However, close to the
critical temperature the renormalization group method will clearly deviate from
mean-field theory. We will here focus on the behavior of the effective two-body
interaction, and defer the discussions concerning the equation of state and
other thermodynamic quantities to Sec.\ IV when we are able to describe also
the condensed phase of the gas.

In the many-body $T-$matrix theory, the chemical potential is renormalized to
$\mu'=\mu-2nT^{2B}({\bf 0,\bf 0};0)$ \cite{USNIST}. This is one of the results
of including all two-body scattering processes, but also incorporating the
effect of the medium on the scattering. Including the latter effect on the
collisions in the gas, which was first carried out explicitly in Refs.\
\cite{STONUCL}, \cite{USVAR} and \cite{USNIST}, is an important step forward in
the correct mean-field treatment of the dilute Bose gas, since including
many-body effects causes the effective interaction to go to zero at the
critical temperature. This resolves a number of problems found in previous
approaches using just the two-body scattering length $a$
\cite{GRIF1,GRIF2,LEEY,STRAL}. With our renormalization group approach we can
corroborate this result and even go somewhat further than that. In our previous
work we included only the many-body effects coming from the ladder diagrams.
However, a class of diagrams that in principle also affects the two-body
interaction are the bubble diagrams. With our current set of renormalization
group equations we can precisely study the effect of these bubble diagrams on
the effective interaction. This is straightforward because we can pinpoint the
ladder and bubble contributions in the equation describing the renormalization
of the interaction. By alternatively including and excluding the bubble
diagrams and then solving the renormalization group equations we can study the
relative
importance of the bubble diagrams on the effective `many-body' scattering
length $a^{eff}$. To avoid any confusion we will adopt the following notation
for the scattering length. The normal `bare' two-body scattering length as
found e.g.\ from analysis of the appropriate association spectra is denoted by
$a$, as usual. The effective scattering length $a^{eff}$ includes also effects
of the medium on two-body scattering, and therefore depends on the specific
approximation used to calculate this effect. Here we consider two such
approximations and denote the corresponding scattering lengths by $a^{MB}$ and
$a^{RG}$. In the first case it is the result of a many-body $T-$matrix
calculation and is defined through $T^{MB}({\bf 0,0,0};0)=4\pi
a^{MB}\hbar^2/m$. In the second case it is the result of a renormalization
group calculation and is defined through $V_{\bf 0}(l=\infty)=4\pi
a^{RG}\hbar^2/m$, irrespective of the fact if bubbles are or are not included
in this calculation. In Fig.\ 4 we depict the ratio of the effective scattering
length resulting from the renormalization group approach to the simple two-body
scattering length $a$ when we include and exclude the bubble diagrams, as a
function of $T/T_c$ and at a density of $1.5\ 10^{12}$ cm$^{-3}$. We conclude
that the effect of the bubbles can be rather large and is in particular
important near the critical temperature. Further away from the critical
temperature the importance of the bubbles rapidly decreases. This implies that
calculating the influence of many-body effects by means of the ladder diagrams
(i.e.\ doing the full many-body $T-$matrix calculation) can quantatively give a
poor estimate for $a^{eff}$. However, qualitatively there is clearly good
agreement as we find from the renormalization group calculation that the
effective scattering length indeed goes to zero at the critical temperature.

As already mentioned before, a bare chemical potential larger than the critical
one, which is positive and yields a flow into the fixed point, corresponds to
an inaccesible density regime. Therefore, we are not able to penetrate the
region with a Bose condensate and cannot describe the Bose gas below the
critical temperature with the renormalization group equations  in Eq.\ (13).
Moreover, the critical exponent $\nu$ found from this set is not a very good
approximation to the true value $\nu=0.67$, even if we would include three-body
interactions. These aspects are intimately related and due to the fact that in
the case of a positive chemical potential, the ${\bf k=0}$ part of the
effective action has a Mexican hat shape. Therefore, we must explicitly break
the symmetry and introduce the condensate density into the action by expanding
the action around its minimum, and not around $\langle\psi\rangle =0$ as was
done in this section. For a negative chemical potential, the above approach is
of course correct as then $\langle\psi\rangle$ is equal to zero.

\section{The symmetry broken phase}
Breaking the symmetry allows us to describe the dilute Bose gas below the
critical temperature. Moreover, also above the critical temperature we find
considerable improvement. This is due to the fact that a positive bare chemical
potential can renormalize to negative values. Thus, starting out in the broken
phase, the fluctuations can restore the symmetry and we end up in the unbroken
phase above the critical temperature. The new set of renormalization group
equations explicitly takes this broken symmetry into account and therefore
gives a much better description than the one resulting from the set of
equations in Eq.\ (13).

\subsection{The flow equations}
In order to expand the action around the correct extremum we have to perform
the shift $a_{{\bf 0},0}\rightarrow a_{{\bf 0},0}+\sqrt{n_0\hbar\beta V}$ which
introduces the condensate density $n_{\bf 0}$. Substituting this in the action
Eq.\ (\ref{basisactie}) leads to the familiar expression \cite{POP1}
\begin{eqnarray}
\label{actiebreak1}
S[a,a^*]&=&-\hbar\beta V(\mu n_{{\bf 0}}-\frac{1}{2}n_{{\bf 0}}^2V_{{\bf
0}})+(-\mu\sqrt{n_{\bf 0}}+n_{\bf 0}\sqrt{n_{\bf 0}}V_{\bf 0})(a^*_{{\bf
0},0}+a_{{\bf 0},0})\nonumber \\
&+&\sum_{{\bf k},n}(-i\hbar\omega_n+\epsilon_{{\bf k}}-\mu+2n_{\bf 0}V_{\bf
0})a^*_{{\bf k},n}a_{{\bf k},n}+\frac{1}{2}n_{\bf 0}V_{\bf 0}\sum_{{\bf
k},n}(a^*_{{\bf k},n}a^*_{{\bf -k},-n}+a_{{\bf k},n}a_{{\bf-k},-n})\nonumber \\
&+&\sqrt{\frac{n_{{\bf 0}}}{\hbar\beta V}}\sum_{\stackrel{\bf k,q}{n,m}}V_{{\bf
0}}(a^*_{{\bf q},m}a^*_{{\bf k-q},n-m}a_{{\bf
k},n}+a^*_{{\bf k+q},n+m}a_{{\bf q},m}a_{{\bf k},n})\nonumber \\
&+&\frac{1}{2}\frac{1}{\hbar\beta V}\sum_{\stackrel{\bf k,k',q}{n,n',m}}V_{{\bf
0}} a^*_{{\bf k+q},n+m}a^*_{{\bf k'-q},n'-m}
a_{{\bf k'},n'}a_{{\bf k},n}~,
\end{eqnarray}
for the action. The magnitude of the condensate is determined by eliminating
the linear term from the action. In first instance we thus find $n_{\bf 0}=\mu/
V_{\bf 0}$. As a result, we can write the action as
\begin{eqnarray}
\label{actiebreakdef}
S[a,a^*]&=&-\hbar\beta V\omega_0+\sum_{{\bf
k},n}(-i\hbar\omega_n+\epsilon_{{\bf k}}+\Gamma_{11})a^*_{{\bf k},n}a_{{\bf
k},n}+\frac{1}{2}\Gamma_{12}\sum_{{\bf k},n}(a^*_{{\bf k},n}a^*_{{\bf
-k},-n}+a_{{\bf k},n}a_{{\bf-k},-n})\nonumber \\
&+&\frac{\Gamma_3}{\sqrt{\hbar\beta V}}\sum_{\stackrel{\bf k,q}{n,m}}(a^*_{{\bf
q},m}a^*_{{\bf k-q},n-m}a_{{\bf
k},n}+a^*_{{\bf k+q},n+m}a_{{\bf q},m}a_{{\bf k},n})\nonumber \\
&+&\frac{1}{2}\frac{V_{\bf 0}}{\hbar\beta V}\sum_{\stackrel{\bf
k,k',q}{n,n',m}}a^*_{{\bf k+q},n+m}a^*_{{\bf k'-q},n'-m}
a_{{\bf k'},n'}a_{{\bf k},n}~,
\end{eqnarray}
introducing $\omega_0=n^2_{\bf 0}V_{\bf 0}/2$ as the lowest order approximation
to the thermodynamic potential density, and defining the vertices
$\Gamma_{11}=\mu-\hbar\Sigma_{11}({\bf 0};0), \Gamma_{12}=\hbar\Sigma_{12}({\bf
0};0)$ and $\Gamma_3$ for which we have in lowest order that
$\Gamma_{11}=\Gamma_{12}=n_{\bf 0}V_{\bf 0}=\mu$ and $\Gamma_3=\sqrt{n_{\bf
0}}V_{\bf 0}=\sqrt{\mu V_{\bf 0}}$. These are the boundary conditions for the
flow equations we derive next.

Deriving the renormalization group equations using Feynman diagrams is in this
case more involved than in the unbroken phase as the number of vertices is
larger, but more so because we now also can have anomalous propagators $\langle
a^*_{{\bf k},n}a^*_{{\bf -k},-n}\rangle$ and $\langle a_{{\bf k},n}a_{{\bf
-k},-n}\rangle$ in these diagrams. The number of diagrams is therefore much
larger and the combinatorics is more complicated. Therefore, we will here use a
different method to obtain the renormalization group equations which does not
explicitly make use of Feynman diagrams, and is therefore very useful and
efficient in this case. It relies on the fact that in Eq.\ (7)
\begin{eqnarray}
\int
d[\psi^*_>]d[\psi_>]\exp\left\{-\frac{1}{\hbar}(S_0[\psi^*_>,\psi_>]+S_I[\psi^*_<, \psi_<,\psi^*_>,\psi_>])\right\} \equiv \exp\left\{-Tr[\ln (-G_>^{-1})] \right\}~,
\end{eqnarray}
where the trace is over Fourier modes in the shell d$\Lambda$ only and $G_>$ is
the Greens function for these fast modes. Taking the shell infinitesimal again,
we can simply calculate ln$(-G_>^{-1})$ because we then only need the part of
the total action which is quadratic in the fast Fourier modes (yielding all
one-loop contributions). The coefficients of this quadratic part also contain
the slow modes because the interaction term $S_I[\psi^*_<,\psi_<,
\psi^*_>,\psi_>]$ couples the slow and fast modes. By simply Taylor expanding
ln$(-G_>^{-1})$ we straightforwardly find the new effective action for the slow
modes and the renormalization group equations of any vertex we would like to
consider.

Thus, we split the fields in slow modes and fast modes as in Eq.\
(\ref{modesplitting}) and find that the part from the action which is quadratic
in the fast modes and only leads to the renormalization of coupling constants
reads
\begin{eqnarray}
\label{fastquadratic}
S^{(2)}[\psi^*_>,\psi_>]&=&{\sum_{{\bf k},n}}'(-i\hbar\omega_n+\epsilon_{{\bf
k}}+\Gamma_{11}+2\Gamma_3(\psi^*_<+\psi_<)+2V_{\bf 0}\mid
\psi_<\mid^2)a^*_{{\bf k},n}a_{{\bf k},n}\nonumber \\
&+&(\frac{1}{2}\Gamma_{12}+\Gamma_3\psi_<+\frac{1}{2}V_{\bf 0}\psi^2_<)
{\sum_{{\bf k},n}}'a^*_{{\bf k},n}a^*_{{\bf -k},-n}\nonumber \\
&+&(\frac{1}{2}\Gamma_{12}+\Gamma_3\psi^*_<+\frac{1}{2}V_{\bf 0}\psi^{*2}_<)
{\sum_{{\bf k},n}}'a_{{\bf k},n}a_{{\bf -k},-n}~,
\end{eqnarray}
where the prime denotes that the sum over momenta is restricted to an
infinitesimal momentum shell d$\Lambda$ at the cutoff. Evaluating the
functional integral over these fast modes leads to adding Tr(ln($-G_>^{-1}$))
to the action for the slow fields, and thus changes the vertices. This quantity
is most easily evaluated by performing a Bogoliubov transformation to
diagonalize Eq.\ (\ref{fastquadratic}) \cite{FETWAL}, and we find that
\begin{eqnarray}
\label{traceresult}
Tr(\ln (-G_>^{-1}))=\frac{\Lambda^2}{2\pi^2}\left[\frac{1}{\beta}\ln
(1-e^{-\beta E_\Lambda})+\frac{1}{2}(E_\Lambda-(\epsilon_\Lambda+\Gamma_{11}+
2\Gamma_3(\psi^*_<+\psi_<)+2V_{\bf 0}\mid \psi_<\mid^2))\right] d\Lambda~,
\end{eqnarray}
where the second term originates from the diagonalization procedure, and the
`dispersion' $E_\Lambda$ is found from
\begin{eqnarray}
\label{Edispersie}
E^2_\Lambda&=&(\epsilon_\Lambda+\Gamma_{11}+2\Gamma_3(\psi^*_<+\psi_<)+2V_{\bf
0}\mid \psi_<\mid^2)^2- \nonumber \\
&&(\Gamma_{12}+\Gamma_3\psi_<+\frac{1}{2}V_{\bf 0}\psi^2_<)(\Gamma_{12}
+\Gamma_3\psi^*_<+\frac{1}{2}V_{\bf 0}\psi^{*2}_<)~.
\end{eqnarray}
In zeroth order in $\psi_<$ and $\psi^*_<$ we retrieve the well-known
Bogoliubov dispersion
$\hbar\omega_\Lambda=\sqrt{(\epsilon_\Lambda+\Gamma_{11})^2 -\Gamma_{12}^2}$,
equal to $\hbar\omega_\Lambda=\sqrt{\epsilon^2_\Lambda
+2\Gamma_{11}\epsilon_\Lambda}$ and thus gapless at $l=0$ due to the equality
$\Gamma_{11}(l=0)= \Gamma_{12}(l=0)$. This corresponds to the Hugenholtz-Pines
theorem \cite{HUPI}. Performing a Taylor expansion in terms of the slow modes
we find the new action. Thus, integrating out a momentum shell renormalizes the
existing vertices, but also generates new terms in the action, and in
particular a linear term. To eliminate this term and remain in the minimum of
the action we again have to perform a small shift in $a_{{\bf 0},0}$. This
implies that the magnitude of the condensate changes as we are integrating out
momentum shells and we also have a flow equation for the condensate density.
Since we omit three-body interactions containing six fields, we also have to
neglect terms containing five fields since they correspond, together with the
condensate field, with a three-body term. As a result, the action remains of
the form written down in Eq.\ (\ref{actiebreakdef}) and the renormalization
group equations can now be obtained for all the vertices of interest. However,
due to the $U(1)-$symmetry of our problem, we can relate some of these vertices
and thereby limit the number of flow equations we actually need to describe the
complete renormalization of the action in Eq.\ (\ref{actiebreakdef}). As this
$U(1)-$symmetry cannot be broken during the process of renormalization, the
action can, at any time, be recast in the explicitly $U(1)$ symmetric form
\begin{eqnarray}
\label{newaction}
S'[\psi^*_<,\psi_<]&=&\int_{0}^{\hbar\beta}d\tau\int d{\bf x}\left(~
\psi^*_<({\bf x},\tau)\left[ \hbar\frac{\partial}{\partial\tau}-
\frac{\hbar^2{\bf \nabla}^2}{2m}-\mu(l)\right] \psi_<({\bf
x},\tau)+\frac{1}{2}V_{\bf 0}(l)\mid\psi_<({\bf x},\tau)\mid^4 \right)~.
\end{eqnarray}
{}From this it is then first of all easy to see that the Hugenholtz-Pines
theorem \cite{HUPI} holds, implying that
$\Gamma_{11}(l)=\Gamma_{12}(l)=\mu(l)$. (See appendix A for an explicit
derivation of this important relation.) Next, also due to the neglect of
three-body interactions, we have that $\Gamma_3(l)=\sqrt{n_{\bf 0}(l)}V_{\bf
0}(l)$ and $V_{\bf 0}(l)=\Gamma_{12}(l)/n_{\bf 0}(l)$. Thus, we only need to
know the flow equations for e.g.\ $n_{\bf 0}(l)$ and $\Gamma_{12}(l)$ and then
the other renormalization group equations can be determined. In the following
we can therefore restrict ourselves to the renormalization of the linear term
and the term proportional to $(\psi^{*2}_<+\psi^2_<)$ as these determine the
change of the condensate density and the anomalous selfenergy $\Gamma_{12}$
respectively. Note that the equality $\Gamma_{11}(l)=\Gamma_{12} (l)$ ensures
that the dispersion $\hbar\omega_\Lambda$ is gapless at any point during
renormalization, as it should.

To arrive at the flow equations for $n_{\bf 0}(l)$ and $\Gamma_{12}(l)$, we
first of all need the linear term $d\Gamma_0(a^*_{{\bf 0},0}+a_{{\bf 0},0})$
that is generated by integrating out a momentum shell. We find that
\begin{eqnarray}
\label{deltaO}
d\Gamma_0=\frac{\Lambda^3}{2\pi^2}\left(\frac{2\Gamma_3(\epsilon_\Lambda+
\Gamma_{11}-\frac{1}{2}\Gamma_{12})}{\hbar\omega_\Lambda}N(\hbar\omega_\Lambda)+ \frac{1}{2}\left(\frac{2\Gamma_3(\epsilon_\Lambda+\Gamma_{11}-\frac{1}{2} \Gamma_{12})}{\hbar\omega_\Lambda}-2\Gamma_3\right)\right)dl~.
\end{eqnarray}
Analogously we find a change $d\Gamma^{(0)}_{12}$ in the anomalous selfenergy.
(See appendix A for details.) However, before we know the full renormalization
of this vertex we have to determine the shift needed to eliminate the linear
term that is generated. Substituting $a_{{\bf 0},0}\rightarrow a_{{\bf
0},0}+\sqrt{\hbar\beta V}d(\sqrt{n_{\bf 0}})$ and retaining only the term
linear in $d(\sqrt{n_{\bf 0}})$ we find that
\begin{eqnarray}
\label{ennulchange}
d(\sqrt{n_{\bf 0}})=-\frac{d\Gamma_0}{\Gamma_{11}+\Gamma_{12}}~
\end{eqnarray}
which influences the renormalization of $\Gamma_{12}$ because we have for the
total renormalization of the anomalous selfenergy
\begin{eqnarray}
\label{gamma12abstract}
d\Gamma_{12}=d\Gamma^{(0)}_{12}+2 \Gamma_3d(\sqrt{n_{\bf 0}})~.
\end{eqnarray}
due to this shift. Using the above mentioned relations between the vertices
implicate from the $U(1)-$symmetry we then make contact with the
renormalization group equations for the unbroken phase by determining the flow
equations for the chemical potential $\mu=\Gamma_{12}$ and the two-body
interaction $V_{\bf 0}=\Gamma_{12}/n_{\bf 0}$. After some algebra we ultimately
find
\begin{mathletters}
\begin{eqnarray}
\label{murgonder}
\frac{d\mu}{dl}=2\mu-\frac{\Lambda^3}{2\pi^2}V_{\bf
0}\left[\frac{2\epsilon_{\Lambda}^3+6\mu\epsilon_{\Lambda}^2+
\mu^3}{2\hbar^3\omega_{\Lambda}^3}(2N(\hbar\omega_{\Lambda})+1)-1
+\frac{\mu(2\epsilon_{\Lambda}+\mu)^2}{\hbar^2\omega_{\Lambda}^2} \beta
N(\hbar\omega_{\Lambda})[N(\hbar\omega_{\Lambda})+1]\right]
\end{eqnarray}
\begin{eqnarray}
\label{vnulrgonder}
\frac{dV_{\bf 0}}{dl}=-V_{\bf 0}-\frac{\Lambda^3}{2\pi^2}V_{\bf 0}^2
\left[\frac{(\epsilon_{\Lambda}-\mu)^2}{2\hbar^3\omega_{\Lambda}^3}
(2N(\hbar\omega_{\Lambda})+1)+\frac{(2\epsilon_{\Lambda}+\mu)^2}
{\hbar^2\omega_{\Lambda}^2} \beta N(\hbar\omega_{\Lambda})
[N(\hbar\omega_{\Lambda})+1]\right]~,
\end{eqnarray}
\end{mathletters}
and the condensate density follows from $n_{\bf 0}(l)=\mu(l)/V_{\bf 0}(l)$.
Breaking the symmetry is irrelevant to the trivial scaling of the vertices, and
thus these are identical to what was found in the previous section. Comparing
these flow equations to Eqs.\ (\ref{murgboven}) and (\ref{vnulrgboven})
omitting $U_{\bf 0}$, we see that both sets coincide when $\mu$ is taken equal
to zero. Thus, the renormalization group equations for negative and positive
chemical potential yield a flow which is continuously differentiable, also at
$\mu=0$.

We now know the renormalization of the chemical potential and the two-body
interaction. Finally, we again have to find the equations describing the
building up of the total and superfluid densities and the thermodynamic
potential as we are integrating out momentum shells. Analogous to the procedure
followed in the previous section we find for the total density, being the sum
of the condensate density and the above condensate density,
\begin{eqnarray}
\label{dichtheidonder}
\frac{dn}{dl}=-\frac{\Lambda^3}{2\pi^2}\left(\frac{\epsilon_\Lambda}{2\hbar
\omega_\Lambda}(2N(\hbar\omega_\Lambda)+1)-\frac{1}{2}\right)e^{-3l}
\end{eqnarray}
with the boundary condition $n(l=0)=n_{\bf 0}(l=0)=\mu (l=0)/V_{\bf 0}(l=0)$.
The superfluid density follows from
\begin{eqnarray}
\label{superdhonder}
\frac{dn_s}{dl}=\frac{dn}{dl}-\frac{\Lambda^3}{3\pi^2}\beta\epsilon_\Lambda
N(\hbar\omega_\Lambda)[N(\hbar\omega_\Lambda)+1]e^{-3l}~,
\end{eqnarray}
and the thermodynamic potential from
\begin{eqnarray}
\label{Omegaonder}
\frac{d\omega}{dl}=\frac{\Lambda^3}{2\pi^2}\left(\frac{1}{\beta}ln(1-e^{-\beta
\hbar\omega_\Lambda})e^{-5l}+\frac{1}{2}(\hbar\omega_\Lambda-\epsilon_\Lambda-
\mu) e^{-3l}\right)~,
\end{eqnarray}
(cf.\ Eq.\ (\ref{traceresult})) with the boundary condition $\omega(l=0)=
\omega_0=n^2_{\bf 0}V_{\bf 0}/2$.

\subsection{The critical temperature of BEC}
With the boundary conditions mentioned above we can again numerically integrate
the renormalization group equations. For a fixed temperature, we vary the value
of the (positive) bare chemical potential, and calculate e.g.\ density and
pressure. The physical effective chemical potential, i.e.\ with the trivial
scaling removed, decreases as we perform the integration, and depending on the
starting value remains positive, renormalizes exactly to zero, or becomes
negative at some value of the integration parameter $l$. In the first case we
start out and remain in the broken phase and are below the critical temperature
of Bose-Einstein condensation, i.e.\ we have a finite condensate density. In
the second case, the condensate density $n_{\bf 0}(l)=\mu(l)/V_{\bf 0}(l)$
renormalizes exactly to zero for $l\rightarrow\infty$, and we are at the
critical conditions for Bose-Einstein condensation. In the third case we
started out in the broken phase, but the fluctuations restore the symmetry. At
the value of $l$ for which the chemical potential becomes negative we have to
continue the integration with the set Eqs.\ (\ref{murgboven}) and
(\ref{vnulrgboven}), and we are thus in the uncondensed phase. Hence, to be
able to describe the dilute Bose gas above, but not too far from the critical
temperature, we need the renormalization group equations for both signs of
$\mu$. In Fig.\ 5 we depict the trajectories resulting from the integration of
Eqs.\ (\ref{murgonder}) and (\ref{vnulrgonder}).

It is evident from this figure that the critical properties of the Bose gas are
determined by this set of equations. By linearizing the flow equations around
the fixed point we can identify the largest eigenvalue $\lambda_+$, and
determine the critical exponent $\nu$ following from this set. We find
$\nu=0.685$, which gives a much better approximation to the critical exponent
than the renormalization group equations studied in Sec.\ III and is to be
compared with the value $\nu=0.67$ found from the $\epsilon -$expansion of the
$O(2)$ model \cite{ZINNJ} and measured in $^4$He experiments. The agreement is
surprisingly good, and together with the fact that we explicitly showed that
three-body effects are negligible, this indicates that we are indeed accurately
describing the Bose gas with the derived renormalization group equations, also
in the critical region. The cause of this good agreement is that although we
only consider the renormalization of $\mu$ and $V_{\bf 0}$, the type of
scattering processes in terms of real (bare) particles we are actually taking
into account are very elaborate. The propagator for the the Bogoliubov
quasiparticles is namely the result of dressing the original bare propagator
$\hbar/(i\hbar\omega_n-\epsilon_\Lambda+\mu)$ with interactions with the
condensate as we use the terms $2n_{\bf 0}V_{\bf 0}a^*_{{\bf k},n}a_{{\bf
k},n}$ and $n_{\bf 0}V_{\bf 0}(a^*_{{\bf k},n}a^*_{{\bf -k},-n}+a_{{\bf
k},n}a_{{\bf -k},-n})/2$ in the zeroth order quadratic part of the action.
Thus, the diagrams we calculate actually contain an infinite number of
scattering processes with the condensate. Therefore, we are describing the
system much better than in Sec.\ III already at this level of renormalization
group.

The first nonuniversal property we concentrate on is the change in the critical
temperature of Bose-Einstein condensation due to interaction effects. This
result is presented also elsewhere \cite{USPRL}, but we will recapitulate it
here because of its experimental significance. At fixed temperatures we vary
the (bare) chemical potential to find the trajectories flowing into the fixed
point. Using Eq.\ (\ref{dichtheidonder}), this yields the critical densities
for these specific temperatures and gives us the $n_c-T$ relation at which
Bose-Einstein condensation occurs. We repeat this for different values of the
scattering length to obtain the dependence of the critical temperature on the
strength of the interaction. In Fig.\ 6 we show the degeneracy parameter
$n_c\Lambda_{th}^3$ found from the renormalization group calculation as a
function of $a/\Lambda_{th}$. As seen from this figure we conclude that the
critical temperature is {\em raised} with respect to the ideal gas value. This
is in qualitative agreement with the recent experiments \cite{rub,nat} and also
preliminary Quantum Monte Carlo calculations seem to confirm this result
\cite{CEPERL}. An indication of an upward shift was also found some time ago by
one of us studying the nucleation of Bose-Einstein condensation \cite{STONUCL}.
{}From our calculations we predict that for the $^{87}$Rb and $^{23}$Na
experiments the critical temperature can be raised with as much as 10 \%, which
appears to be a very promising result because one might expect that an effect
of this magnitude can very well be measured in future, more accurate,
experiments. It is important to note that this shift in $n_c\Lambda^3_{th}$ can
be observed in magnetically trapped atomic gases if one directly measures the
density in the center of the trap at the critical temperature. One should in
particular {\em not} measure the total number of particles, because this
involves the density profile in the trap and due to the repulsive nature of the
interactions thus tends to obscure the effect \cite{GIORSTRI}.

The reason for a higher critical temperature, or more precisely, a lower
critical density is the following. The effective chemical potential
renormalizes from a positive initial value exactly to zero. Consequently, we
have the Bogoliubov dispersion in the equation for the density, and this
depresses the occupation of the non-zero momentum states relative to the ideal
gas case, where we would just have $\epsilon_\Lambda$ in the Bose-Einstein
distribution function. The magnitude of the effect is related to the behavior
of the chemical potential when renormalizing to the fixed point value
$\mu^*=\epsilon_\Lambda$. Suppose we effectively have $\mu(l)=\alpha
\epsilon_\Lambda$ independent of $l$, with some positive $\alpha$ smaller than
1. We can then translate the differential equation for the density into an
ordinary integral over momentum space following the inverse procedure from
which we found the flow equations in Sec.\ III. Doing so, we find that the
Bogoliubov dispersion effectively becomes $\hbar\omega_{\bf
k}=\sqrt{1+2\alpha}\epsilon_{\bf k}$. Thus, this essentially boils down to a
mass renormalization, and the change in the critical temperature is directly
related to the magnitude of this renormalization. Starting with the equation
describing the building up of the above condensate density (cf.\ Eq.\
(\ref{dgam11A}) from appendix A) one can easily show that we approximately have
$n_c\Lambda^3_{th}=(1+\alpha)g_{3/2}(1)/ (1+2\alpha)$. For an increase of 10 \%
in the critical temperature, we find from this result that $\alpha$ must be
equal to 0.2. Indeed, this turns out to be the typical value of $\alpha$ in the
thermal regime where the contributions to the density are largest.

\subsection{The region $na\Lambda^2_{th}\ll 1$}
To obtain also information on the properties of the Bose gas below the critical
temperature we have to start with a chemical potential which always remains
positive under renormalization. We are then always in the condensed phase and
have to use the renormalization group equations (\ref{murgonder}) and
(\ref{vnulrgonder}). Written as ordinary one-loop integrals, these equations
contain infrared divergencies and a straightforward perturbative analysis is
not possible. This is a well known problem in the theory of the interacting
Bose gas \cite{GRIFB,NEPOMN}. However, doing the calculation by means of the
renormalization group approach, this problem is in principle resolved due to
the resummation which is automatically performed. The presence of infrared
divergencies causes both the physical chemical potential and two-body
interaction in the renormalization group approach to renormalize to zero,
instead of becoming infinite as they would in a regular one-loop calculation.
Indeed, one can show from Eqs.\ (\ref{murgonder}) and (\ref{vnulrgonder}) that
for $l\rightarrow \infty$ we have $\mu\propto e^{l}$ and $V_{\bf 0}\propto
e^{-2l}$. These scalings are different from the trivial scalings $\mu\propto
e^{2l}$ and $V_{\bf 0}\propto e^{-l}$, and are therefore termed anomalous.
Physically, the anomalous scaling implies an effective energy and momentum
dependence of the coupling constants, as we show explicitly in appendix B. In
particular, the anomalous scaling we find leads to the result that the chemical
potential, and thus the normal and anomalous selfenergies, as well as the
two-body interaction, behave linearly with $k$ for low momenta, i.e.\ our
renormalization group approach reveals that $\hbar\Sigma_{11}({\bf k};0)\propto
k$ and $\hbar\Sigma_{12}({\bf k};0)\propto k$. This behavior, implying that
$\hbar\Sigma_{11}({\bf 0};0)=\hbar\Sigma_{12} ({\bf 0};0)=0$, is an exact
result
for low momenta \cite{NEPOMN} which we explicitly recover here. The
consequences for the application of renormalization group are however twofold.
First, it implies that the Bogoliubov dispersion $\hbar\omega_{\bf k}$ does not
possess a sound mode anymore since the low momentum behavior is not linear, but
instead we have $\hbar\omega_{\bf k}\propto k^{3/2}$. This is of course an
incorrect result \cite{GRIFB} indicating that Eqs.\ (\ref{murgonder}) and
(\ref{vnulrgonder}) are insufficient to describe the sound mode. Indeed,
following the argument in Sec.\ III we neglected the renormalization of the
(marginal) $\partial/\partial\tau$ and $\mid\nabla\mid^2$ terms since the
interactions were anticipated to be momentum and energy independent. However,
below the critical temperature we see that the momentum and energy dependence
of the selfenergies does become important, and should therefore be included in
the renormalization group calculation by Taylor expanding the selfenergies in
terms of the external momentum and frequency. Put differently, the anomalous
dimension $\eta$ is no longer small below the critical temperature. The extra
renormalization group equations obtained in this manner may change the
particular anomalous scaling found above, but since $\hbar\Sigma_{11}({\bf
0};0)=\hbar\Sigma_{12}({\bf 0};0)=0$ is an exact result we still expect to have
anomalous scaling of the coupling constants. The effect of the extra
renormalization group equations will be to change the dispersion relation in
such a way that the linear sound mode is recovered. We will not pursue this
calculation here, but postpone it to future work.

The reason for the anomalous scaling and consequently the disappearance of the
sound mode, is caused by the infrared divergence in the one-loop expressions
for the selfenergies and two-body interaction. This can be traced back to the
behavior of the coherence factors $u_{\bf k}$ and $v_{\bf k}$ of the Bogoliubov
transformation diagonalizing the quadratic part of the action. The Bogoliubov
transformation is given by
\begin{mathletters}
\label{bogtrafo}
\begin{eqnarray}
a_{{\bf k},n}&=&u_{\bf k}b_{{\bf k},n}-v_{\bf k}b^*_{{\bf -k},-n} ,
\end{eqnarray}
\begin{eqnarray}
a^*_{{\bf k},n}&=&u_{\bf k}b^*_{{\bf k},n}-v_{\bf k}b_{{\bf -k},-n}~,
\end{eqnarray}
\end{mathletters}
where $b^*_{{\bf k},n}$ and $b_{{\bf k},n}$ are the Fourier components of the
fields creating respectively annihilating a Bogoliubov quasiparticle. The
coherence factors are given by
\begin{mathletters}
\label{coherence}
\begin{eqnarray}
u_{\bf k}&=&\frac{1}{2}\left(\sqrt{\frac{\hbar\omega_{\bf k}}{\epsilon_{\bf
k}}}+\sqrt{\frac{\epsilon_{\bf k}}{\hbar\omega_{\bf k}}}\right) ,
\end{eqnarray}
\begin{eqnarray}
v_{\bf k}&=&\frac{1}{2}\left(\sqrt{\frac{\hbar\omega_{\bf k}}{\epsilon_{\bf
k}}}-\sqrt{\frac{\epsilon_{\bf k}}{\hbar\omega_{\bf k}}}\right)~,
\end{eqnarray}
\end{mathletters}
and satisfy the requirement that $u^2_{\bf k}- v^2_{\bf k}=1$ because the
Bogoliubov transformation is unitary. As the Bogoliubov dispersion is given by
$\hbar\omega_{\bf k}=\sqrt{\epsilon^2_{\bf k}+2\mu\epsilon_{\bf k}}$, it is
easy to see that $u_{\bf k}\rightarrow 1$ and $v_{\bf k}\rightarrow 0$ for
large momenta, and both factors behave as $1/\sqrt{k}$ as $k$ tends to zero.
The region where the crossover occurs between these two regimes is determined
by the parameter $na\Lambda^2_{th}$, found from comparing the chemical
potential with the thermal energy, and also determining the crossover from
quadratic to linear behavior of the Bogoliubov dispersion. When
$na\Lambda^2_{th}\gg 1$ the linear regime of the dispersion is extremely
important in determining the properties of the Bose gas. Conversely, when
$na\Lambda^2_{th}\ll 1$, the Bogoliubov dispersion can be approximated by the
normal dispersion $\epsilon_{\bf k}+\mu$, except for a very small region at low
momenta, which is then not very important and can for most practical purposes
be neglected. In the following, we therefore concentrate on the regime
$na\Lambda^2_{th}\ll 1$, covering a large temperature interval below the
critical temperature, and use $u_{\bf k}=1$ and $v_{\bf k}=0$ in contributions
to the renormalization of the vertices that contain infrared divergencies.
Otherwise, we will use the full expression Eq.\ (\ref{coherence}). This
procedure turns out to be necessary to ensure the Hugenholtz-Pines theorem to
be satisfied, as is shown in appendix A.

In the method used in the previous section it is not possible to pinpoint the
Bogoliubov coherence factors at any stage of the derivation of the
renormalization group equations. Therefore, we really have to go through the
calculation of the Feynman diagrams relevant to the various vertices we are
interested in. In our case these are the linear term of the action and the
anomalous selfenergy. The one-loop diagrams of interest are depicted in Fig.\ 7
and contain also the anomalous propagators $\langle a^*_{{\bf k},n}a^*_{{\bf
-k},-n}\rangle$ and $\langle a_{{\bf k},n}a_{{\bf -k},-n}\rangle$. Using the
Bogoliubov transformation, these diagrams are straightforward to calculate, and
we find that for the renormalization of the condensate density both diagrams
for $\Gamma_0$ contribute, but that for the renormalization of the anomalous
selfenergy only diagrams $A$ and $B$ give contributions that do not contain
infrared divergencies. The explicit calculation of these diagrams is presented
in appendix A. Using the expressions obtained there, we find that the flow
equations become
\begin{mathletters}
\begin{eqnarray}
\label{muzonderir}
\frac{d\mu}{dl}=2\mu-\frac{\Lambda^3}{2\pi^2}V_{\bf 0}\left(\frac{\epsilon_
\Lambda+\mu}{\hbar\omega_\Lambda}(2N(\hbar\omega_\Lambda)+1)-1+4\beta\mu
N(\hbar\omega_\Lambda)[N(\hbar\omega_\Lambda)+1]\right)~,
\end{eqnarray}
\begin{eqnarray}
\label{vnulzonderir}
\frac{dV_{\bf 0}}{dl}=-V_{\bf 0}-\frac{\Lambda^3}{2\pi^2}V^2_{\bf 0}\left(
\frac{1+2N(\hbar\omega_\Lambda)}{2\hbar\omega_\Lambda}+4\beta N(\hbar\omega_
\Lambda)[N(\hbar\omega_\Lambda)+1]\right)~,
\end{eqnarray}
\end{mathletters}
where we again made use of the relations $\mu(l)=\Gamma_{11}(l)=\Gamma_{12}(l)=
\sqrt{n_{\bf 0}(l)}\Gamma_3(l)=n_{\bf 0}(l)V_{\bf 0}(l)$ due to
$U(1)-$symmetry. Note that also this set coincides with Eqs.\ (\ref{murgboven})
and (\ref{vnulrgboven}) when the chemical potential is equal to zero, so the
flow is also in this case everywhere continuous and continuously
differentiable. Moreover, the equations for the density, superfluid density and
thermodynamic potential do not contain an infrared divergency, so the flow
equations for these quantities remain the same and are given by Eqs.\
(\ref{dichtheidonder}), (\ref{superdhonder}) and (\ref{Omegaonder}),
respectively.

\subsection{Analysis of the flow equations}
At this point we can describe any point in the phase diagram of the dilute Bose
gas. For negative chemical potential we have Eqs.\ (\ref{murgboven}) and
(\ref{vnulrgboven}), for positive chemical potential we must use Eqs.\
(\ref{muzonderir}) and (\ref{vnulzonderir}). Moreover, we have to combine both
sets when we are not too far below the critical density when the chemical
potential changes sign during application of the renormalization group
transformation. Using Eq.\ (37), we find that the critical temperature of
Bose-Einstein condensation changes with less than 0.1 \% compared to the more
accurate result found using Eq.\ (31). Therefore, this shows that using the set
not containing the infrared divergencies essentially leads to the same results,
implying that it is indeed correct to neglect the linear part of the Bogoliubov
dispersion. Having proven this explicitly, we will in the following present the
results of the renormalization group calculation of the effective two-body
interaction, i.e.\ the many-body scattering length $a^{eff}$, the condensate
and superfluid densities and the $p-n^{-1}-$diagram below as well as above the
critical temperature, and compare them in all cases with the results from the
many-body $T-$matrix calculation.

We start with the scattering length $a^{eff}$. Above the critical temperature
it is straightforward to take into account only the contributions of the ladder
diagrams, or to include the effect of the bubble diagrams as well. Below the
critical temperature there are no clear ladder and bubble diagrams, so the
various contributions are in principle not clearly associated with ladders or
bubbles. However, comparing the differential equations for positive and
negative $\mu$ we can conclude that the first nontrivial term on the righthand
side of Eq.\ (\ref{vnulzonderir}) is effectively related to a ladder
contribution, and that the second nontrivial term is effectively related to a
bubble diagram. We depict in Fig.\ 8 the scattering length $a^{eff}$,
normalized to the two-body scattering length $a=52a_0$ for $^{23}$Na, when
including and excluding the bubble contributions ($a^{RG}/a$), and also the
result found from the many-body $T-$matrix calculation ($a^{MB}/a$)
\cite{USVAR}. As our approach is restricted to the regime $na\Lambda^2_{th}\ll
1$, we present no results for $na\Lambda^2_{th}$ larger than one. The
renormalization of the effective two-body interaction to zero at the critical
temperature, a result already found in the many-body $T-$matrix approach, turns
out to be correct. Indeed, this can be easily understood from the
renormalization group equations. A fixed point $(\mu^*,V^*_{\bf 0})$ is present
in the set $\{\mu(l),V_{\bf 0}(l)e^{2l}\}$, which means that for
$l\rightarrow\infty$ the physical two-body interaction, i.e.\ with the trivial
scaling removed, behaves as $V^*_{\bf 0}e^{-l}$ and thus goes to zero. The
depression in the scattering length around the critical temperature occurs in a
fairly large temperature interval. Note furthermore that applying the
renormalization group equations (\ref{murgonder}) and (\ref{vnulrgonder}) would
lead to $a^{RG}=0$ everywhere below the critical temperature due to the
anomalous scaling found from this set. This property is actually expected to
hold true even in a more elaborate renormalization group calculation and is an
issue worthwile studying as it may have important consequences for e.g.\ the
exact form of the Gross-Pitaevskii equation describing the condensate.

Next, we turn to the equation of state. For the ideal Bose gas we have in
general that
\begin{eqnarray}
\label{idealgas}
n\Lambda_{th}^3=n_{\bf 0}\Lambda_{th}^3+g_{3/2}(\zeta)
\end{eqnarray}
where $g_n(z)$ is a Bose-function defined as
\begin{eqnarray}
\label{gndef}
g_n(z)=\frac{1}{\Gamma (n)}\int_0^\infty\frac{y^{n-1}}{z^{-1}e^y-1}dy
\end{eqnarray}
and $\zeta=e^{\beta\mu}$ the fugacity. In the many-body $T-$matrix theory above
the critical temperature we have that the chemical potential is renormalized to
$\mu'=\mu-2nT^{2B}({\bf 0,\bf 0};0)$ \cite{USNIST}. Below the critical
temperature the dispersion changes to the Bogoliubov dispersion and the
equation for the density is in essence given by Eq. (\ref{dichtheidonder}),
recast in the
form of an ordinary one-loop integral. In Fig.\ 9 we plot $n\Lambda_{th}^3$ for
the ideal Bose gas, following from the many-body $T-$matrix calculation, and
from the renormalization group calculation. In the latter case, we numerically
integrate the flow equations for a fixed temperature, and vary the value of the
bare chemical potential. This changes the total density in the system, and
therefore the value of $n\Lambda^3_{th}$. For the ideal Bose gas, positive
values of the chemical potential are not allowed. With the repulsive
interactions taken into account, a positive value is possible and the fugacity
can be larger than one. The renormalization group calculation yields the same
density for a chemical potential slightly above and slightly below the critical
chemical potential. This double valuedness in the density occurs in an
extremely small region around the critical density and the effect is smaller
than 1 promille in the case of $^{23}$Na. For $^1$H, with a scattering length
$a=1.34 a_0$, the effect is even far below the promille level. However, it is
important to note that this effect is much smaller than the change in density
we obtain when we include three-body interactions, being approximately 1 \%
(cf. Sec.\ III). Therefore, the double valuedness we find cannot be trusted
physically and should be neglected. It presumably disappears when we include
three-body effects or extend the renormalization group calculation otherwise.

When the bare chemical potential decreases the gas becomes more and more dilute
and the influence of the interactions can better and better be accounted for
using mean-field theory. This fact is actually evident from the renormalization
group equations in Eq.\ (13). When the chemical potential is large and
negative, the Bose-factors $N(\epsilon_\Lambda-\mu)$ are strongly depressed and
there is hardly any many-body effect on the renormalization of the two-body
interaction so we will find that it is just renormalized to $T^{2B}({\bf 0,\bf
0};0)=4\pi a\hbar^2/m$. Indeed, the same is true in the many-body $T-$matrix
calculation \cite{USVAR}, since we have that $T^{MB}({\bf 0,\bf 0,\bf
0};0)\approx T^{2B}({\bf 0,\bf 0};0)$ when the system is extremely dilute,
i.e.\ there is no effect of the medium in this regime. In addition, the
differential equations for $\mu$ and $V_{\bf 0}$ are now almost decoupled, and
consequently the chemical potential will renormalize approximately to
$\mu-2nT^{2B}({\bf 0},{\bf 0};0)$, which is only a small effect in this limit.
Therefore, the renormalization group equations can practically be recast in the
regular one-loop expressions one encounters in the many-body $T-$matrix theory
and the results we find in this regime are approximately the same.

In Figs.\ 10(a) and 10(b) we depict the condensate and superfluid fractions as
a function of temperature for a density of 1.5\ 10$^{12}$ cm$^{-3}$ both from
the renormalization group and many-body $T-$matrix calculation. Again, not too
close to the critical temperature we have good agreement. Note that the
superfluid density has not yet become zero at the temperature where the
condensate density vanishes. To explain this aspect we will focus on the
equation for the superfluid density in the unbroken phase. When we interpret
the fact that the chemical potential is renormalized as we integrate out
momentum shells as a chemical potential depending on momentum, we can express
the superfluid density as a regular integral over momentum space, i.e.
\begin{eqnarray}
\label{superfluid2}
n_s=\int\frac{d^3{\bf k}}{(2\pi)^3}\frac{1}{e^{\beta(\epsilon_{\bf
k}-\mu(k))}-1}-\int\frac{d^3{\bf k}}{(2\pi)^3}\frac{2}{3} \beta\epsilon_{\bf
k}\frac{e^{\beta(\epsilon_{\bf k}-\mu(k))}}{(e^{\beta(\epsilon_{\bf
k}-\mu(k))}-1)^2}~.
\end{eqnarray}
Writing $\mu(k)=\mu_0+\delta\mu(k)$ and performing a Taylor expansion, we find
to first order in $\delta\mu$ that
\begin{eqnarray}
\label{superfluidrg}
n_s=\int\frac{d^3{\bf k}}{(2\pi)^3}\beta N(\epsilon_{\bf k}-\mu_0)
[N(\epsilon_{\bf k}-\mu_0)+1][1-\frac{2}{3}\beta\epsilon_{\bf
k}(2N(\epsilon_{\bf k}-\mu_0)+1)]~\delta\mu (k) ,
\end{eqnarray}
where the lowest order terms drop out. This is due to the fact that these terms
yield the expression for the superfluid density in an ideal Bose gas above the
critical temperature which can be shown to be exactly equal to zero.  It is
clear from Eq.\ (\ref{superfluidrg}) that the superfluid density will in
general not be equal to zero. The behavior of $\delta\mu (k)$ would have to be
very special to give a superfluid density exactly equal to zero. It is
therefore not surprising that we find from our limited set of renormalization
group equations a different temperature for which the condensate and superfluid
densities vanish. For the situation depicted in Fig.\ 10 we have however that
$\Delta T/T_c$ is only about $8\ 10^{-3}$. Extending the renormalization group
calculation would in principle lead to a superfluid density which vanishes at
the same temperature as where the condensate density becomes zero.

Finally, we also present the pressure of the dilute Bose gas as a function of
the inverse density. Using Eq.\ (\ref{Omegaonder}), we depict this behavior in
Fig.\ 11 for a $^{23}$Na gas at 1 $\mu$K, together with the result of the
many-body $T-$matrix calculation. Although the critical densities are
different, we find a fairly good agreement between the two curves and the
difference between the renormalization group calculation and the many-body
$T-$matrix calculation is small when it concerns the pressure of the gas.
However, concerning the other nonuniversal properties discussed in this
article, the difference can be substantial near the critical temperature as we
have seen.

\section{Conclusions}
In summary, we have derived the renormalization group equations for a dilute
Bose gas. To obtain these equations we have to make a distinction between the
case of a negative and a positive chemical potential. In the latter case we
have to take the presence of a condensate into account, and the renormalization
group equations are different from the ones valid in the unbroken phase. Our
philosophy of using the renormalization group approach is not to study
universal properties of the dilute Bose gas, which are the same as for the
$O(2)-$model in three dimensions, but to make quantative predictions about
various nonuniversal properties of this system. To achieve this goal one has to
find a method to eliminate the ultraviolet cutoff dependence inherent to the
application of the renormalization group. Our knowledge about the two-body
scattering problem is sufficient in this respect and we can fix the
renormalization group equations in this manner. We compared results of the
renormalization group with mean-field calculations and showed the difference to
vanish in appropriate limits. We also checked the influence of three-body
effects, which turned out to be unimportant even in the critical region.
Furthermore, we showed that the influence of bubble diagrams on the effective
interaction can be fairly large, and that the effective scattering length
$a^{eff}$ vanishes when approaching the critical temperature. This confirms
earlier results \cite{USPRL,USNIST} and is of importance for a lot of current
work concerning condensate properties in trapped Bose gases. It implies that
there can in principle be an important change in the results of calculations
that make use of the nonlinear Schr\"{o}dinger equation when determining these
properties. For that purpose the effective scattering length $a^{eff}$ is of
importance, and not the two-body scattering length $a$.

Next, we derived the renormalization group equations in the broken phase, i.e.\
for positive chemical potential, and showed that the transition to the
Bose-Einstein condensed phase is contained in this set of renormalization group
equations. We found a critical exponent $\nu=0.685$, which agrees very well
with $\nu=0.67$ found in studies of the $O(2)-$model, and furthermore
calculated the effect of the interactions on the critical temperature of the
phase transition. The change can be as much as 10 \% in the current type of
experiments and may be measured if one can improve the precision in determining
the temperature and the central density in the trap.

As these latter renormalization group equations lead to a dispersion relation
which has no linear part, i.e.\ there is no sound mode, we have to restrict
ourselves to the regime $na\Lambda^2_{th}\ll 1$, where the linear part of the
dispersion is unimportant for determining properties such as density and
pressure. The region where $na\Lambda^2_{th}\ll 1$ is currently still the most
interesting one from the experimental point of view although the other region
is certainly within reach. Including the renormalization of the time derivative
and gradient terms in the action is expected to resolve the problem of the
disappearing of the sound mode, but we postponed this to future work. A
resolution of this problem would correspond to a resolution of the long
standing problem of the infrared divergencies in the perturbation expansion
around the Bogoliubov theory.

We concluded this work with using the renormalization group to calculate
superfluid and condensate densities as a function of temperature and also a
$p-n^{-1}-$diagram is presented. Of course the renormalization group can be
used to find out many more things about the dilute Bose gas. Note e.g.\ that it
is in principle also possible to use the renormalization group method to find
the quantative form of the correlation function as it is the Fourier transform
of the occupation number $N_{\bf k}$. This can also be translated into a
differential equation as in Eq.\ (19) where we now have the distance $r$ as a
free parameter. Also the specific heat may be calculated. In principle, our
results pertain to a homogeneous Bose gas, but in situations where the
application of a local density approximation is allowed, they are also
applicable to trapped Bose gases. Moreover, we indicated that it is in
principle also possible to set up a renormalization group calculation for the
inhomogeneous case.

Finally, we want to note that the procedure of renormalization group as
described in this article can in principle also be used to study the
Kosterlitz-Thouless transition to the superfluid phase in a two-dimensional
Bose gas. However, there are difficulties in this case connected with the fact
that all coupling constants are relevant at the critical temperature.
Nonetheless, work along these lines is in progress since a number of
experiments are currently under construction which aim at reaching the
Kosterlitz-Thouless phase in doubly spin-polarized atomic hydrogen adsorbed on
a superfluid helium film.

\section*{Acknowledgments}
We acknowledge helpful discussions with Eric Cornell, Wolfgang Ketterle and
Steve Girvin.

\section*{Appendix A}
\setcounter{equation}{0}
\renewcommand{\theequation}{A.\arabic{equation}}
In this appendix we go through some of the technicalities of the calculation of
the one-loop Feynman diagrams for the coupling constants $\Gamma_{11}$ and
$\Gamma_{12}$, i.e.\ essentially the normal and anomalous selfenergies, and
explicitly show the Hugenholtz-Pines theorem to hold in our renormalization
group approach. This theorem states that \cite{HUPI}
\begin{eqnarray}
\label{hugenhpines}
\mu=\hbar\Sigma_{11}({\bf 0};0)-\hbar\Sigma_{12}({\bf 0};0)~,
\end{eqnarray}
where $\hbar\Sigma_{11}({\bf 0};0)$ and $\hbar\Sigma_{12}({\bf 0};0)$ are the
irreducible normal and anomalous selfenergies respectively. In our notation
this relation reads
\begin{eqnarray}
\label{onshupi}
\Gamma_{11}=\Gamma_{12}~.
\end{eqnarray}
Fig.\ 7 in the main text contains the diagrams renormalizing $\Gamma_{12}$ and
in Fig.\ 12 we depict the one-loop diagrams renormalizing $\Gamma_{11}$. Using
the designation of the diagrams as in these figures one can show, after going
through the combinatorics, that the (infinitesimal) change of the vertices
after integrating out an infinitesimal momentum shell is given by
\begin{mathletters}
\renewcommand{\theequation}{A.\arabic{equation}a}
\begin{eqnarray}
\label{gam11diagram}
d\Gamma_{11}=4d\Gamma^A_{11}+4d\Gamma^B_{11}+2d\Gamma^C_{11}+4d\Gamma^D_{11}
+4d\Gamma^E_{11}+4d\Gamma^F_{11}+4\Gamma_3 d(\sqrt{n_{\bf 0}})~,
\end{eqnarray}
\renewcommand{\theequation}{A.\arabic{equation}b}
\begin{eqnarray}
\label{gam12diagram}
d\Gamma_{12}=2d\Gamma^A_{12}+4d\Gamma^B_{12}+4d\Gamma^C_{12}+4d\Gamma^D_{12}
+4d\Gamma^E_{12}+2d\Gamma^F_{12}+2\Gamma_3 d(\sqrt{n_{\bf 0}})~.
\end{eqnarray}
\end{mathletters}
The last term in both expressions originates from the shift in $a_{{\bf 0},0}$
required to eliminate the linear term d$\Gamma_0 (a^*_{{\bf 0},0}+a_{{\bf
0},0})$ from the action. From Figs.\ 7 and 12 it is clear that a number of
diagrams contributing to $\Gamma_{11}$ or $\Gamma_{12}$ are mathematically
identical. Most notably we have $d\Gamma^B_{11}=d\Gamma^F
_{11}=d\Gamma^D_{12}=d\Gamma^E_{12}$, $d\Gamma^D_{11}=d\Gamma^C_{12}=
d\Gamma^F_{12}$, and $d\Gamma^E_{11}=d\Gamma^B_{12}$. Therefore, there are only
six independent diagrams, namely $d\Gamma^A_{11}, d\Gamma^B_{11},
d\Gamma^C_{11}, d\Gamma^D_{11}, d\Gamma^E_{11}$, and $d\Gamma^A_{12}$. Using
the Bogoliubov transformation from Eq.\ (\ref{bogtrafo}), it follows that
\begin{eqnarray}
\label{verwwaardena}
\langle a^*_{{\bf k},n}a_{{\bf k},n}\rangle=u^2_{\bf k}\langle b^*_{{\bf
k},n}b_{{\bf k},n}\rangle+v^2_{\bf k}\langle b^*_{{-\bf k},-n}b_{{-\bf
k},-n}\rangle~,
\end{eqnarray}
\begin{eqnarray}
\label{verwwaardenb}
\langle a^*_{{\bf k},n}a^*_{{-\bf k},-n}\rangle=\langle a_{{\bf k},n}a_{{-\bf
k},-n}\rangle=-u_{\bf k}v_{\bf k}(\langle b^*_{{\bf k},n}b_{{\bf k},n}\rangle
+\langle b^*_{{-\bf k},-n}b_{{-\bf k},-n}\rangle)~,
\end{eqnarray}
with $u_{\bf k}$ and $v_{\bf k}$ given in Eq.\ (\ref{coherence}), and these
diagrams are now straightforward to calculate by applying the usual Feynman
rules \cite{FETWAL}. We have
\begin{mathletters}
\renewcommand{\theequation}{A.\arabic{equation}a}
\begin{eqnarray}
\label{dgam11A}
d\Gamma^A_{11}=\frac{1}{2}V_{\bf 0}dn'=\frac{\Lambda^3}{4\pi^2}V_{\bf 0}\left(
\frac{1}{2}(u^2_\Lambda+v^2_\Lambda)(2N(\hbar\omega_\Lambda)+1)-\frac{1}{2}
\right)e^{-3l}dl ,
\end{eqnarray}
\renewcommand{\theequation}{A.\arabic{equation}b}
\begin{eqnarray}
\label{dgam11B}
d\Gamma^B_{11}=\frac{\Lambda^3}{2\pi^2}\Gamma^2_3(u^3_\Lambda v_\Lambda+
u_\Lambda v^3_\Lambda)\left(\frac{1+2N(\hbar\omega_\Lambda)}{2\hbar\omega_
\Lambda}+\beta N(\hbar\omega_\Lambda)[N(\hbar\omega_\Lambda)+1]\right)
e^{-3l}dl~,
\end{eqnarray}
\renewcommand{\theequation}{A.\arabic{equation}c}
\begin{eqnarray}
\label{dgam11C}
d\Gamma^C_{11}=-\frac{\Lambda^3}{2\pi^2}\Gamma^2_3\left( (u^4_\Lambda+
v^4_\Lambda)\frac{1+2N(\hbar\omega_\Lambda)}{2\hbar\omega_\Lambda}+2u^2_\Lambda
v^2_\Lambda\beta N(\hbar\omega_\Lambda)[N(\hbar\omega_\Lambda)+1]\right)
e^{-3l}dl ,
\end{eqnarray}
\renewcommand{\theequation}{A.\arabic{equation}d}
\begin{eqnarray}
\label{dgam11D}
d\Gamma^D_{11}=-\frac{\Lambda^3}{\pi^2}\Gamma^2_3 u^2_\Lambda v^2_\Lambda
\left(\frac{1+2N(\hbar\omega_\Lambda)}{2\hbar\omega_\Lambda}+ \beta
N(\hbar\omega_\Lambda)[N(\hbar\omega_\Lambda)+1]\right)e^{-3l}dl~,
\end{eqnarray}
\renewcommand{\theequation}{A.\arabic{equation}e}
\begin{eqnarray}
\label{dgam11E}
d\Gamma^E_{11}=-\frac{\Lambda^3}{2\pi^2}\Gamma^2_3 \left(2u^2_\Lambda v^2_
\Lambda \frac{1+2N(\hbar\omega_\Lambda)}{2\hbar\omega_\Lambda}+(u^4_\Lambda +
v^4_\Lambda)\beta N(\hbar\omega_\Lambda)[N(\hbar\omega_\Lambda)+1]\right)
e^{-3l}dl~,
\end{eqnarray}
\renewcommand{\theequation}{A.\arabic{equation}f}
\begin{eqnarray}
\label{dgam12A}
d\Gamma^A_{12}=\frac{1}{2}V_{\bf 0}d\tilde{n}=-\frac{\Lambda^3}{4\pi^2}V_{\bf
0}u_\Lambda v_\Lambda(1+2N(\hbar\omega_\Lambda))e^{-3l}dl~,
\end{eqnarray}
\end{mathletters}
introducing $dn'=\Lambda^3\sum_n\langle a^*_{\Lambda,n} a_{\Lambda,n}\rangle
e^{-3l}dl /2\pi^2$ and $d\tilde{n}=\Lambda^3\sum_n\langle
a^*_{\Lambda,n}a^*_{-\Lambda, -n}\rangle e^{-3l}dl/2\pi^2$ with which one can
easily verify that
\begin{eqnarray}
\label{dnnulap}
d(\sqrt{n_{\bf 0}})=-\frac{d\Gamma_0}{\Gamma_{11}+\Gamma_{12}}
=-\frac{1}{\Gamma_{11}+\Gamma_{12}}\Gamma_3(2dn'+d\tilde{n})~.
\end{eqnarray}
Combining these equations leads to the flow equations for $\Gamma_{11}$ and
$\Gamma_{12}$. However, we here concentrate on the Hugenholtz-Pines relation,
which is satisfied at $l=0$, and therefore remains valid if also
$d\Gamma_{11}=d\Gamma_{12}$ at $l=0$. Using the above mentioned equalities of
the various diagrams, this equality reduces in first instance to
\begin{eqnarray}
\label{hupistep1}
2d\Gamma^A_{11}+d\Gamma^C_{11}+\Gamma_3 d(\sqrt{n_{\bf 0}})=d\Gamma^A_{12}
+d\Gamma^C_{12}=d\Gamma^A_{12}+d\Gamma^D_{11}~.
\end{eqnarray}
However, due to the $U(1)-$symmetry of the action $S[\psi^*,\psi]$ we have at
any value of $l$ that $\Gamma^2_3=n_{\bf 0}V^2_{\bf 0}=\Gamma_{12}V_{\bf 0}$
and at $l=0$, when $\Gamma_{11}=\Gamma_{12}$, Eq.\ (\ref{hupistep1}) reduces
with the help of Eq.\ (\ref{dnnulap}) to
\begin{eqnarray}
\label{hupistep2}
d\Gamma^C_{11}-d\Gamma^D_{11}=V_{\bf 0}d\tilde{n}~,
\end{eqnarray}
or equivalently
\begin{eqnarray}
\label{hupistep3}
-\frac{\Lambda^3}{2\pi^2}\Gamma^2_3(u^4_\Lambda-2u^2_\Lambda v^2_\Lambda+
v^4_\Lambda)\frac{1+2N(\hbar\omega_\Lambda)}{2\hbar\omega_\Lambda}=
-\frac{\Lambda^3}{2\pi^2}V_{\bf 0}u_\Lambda v_\Lambda(1+2N(\hbar\omega_
\Lambda))~.
\end{eqnarray}
Note that, due to the fact that $u^2_\Lambda-v^2_\Lambda=1$, the left-hand side
of this equation actually does not contain Bogoliubov coherence factors.
{}From the definition of the coherence factors in Eq.\ (\ref{coherence}) it is
now easy to show that indeed, at $l=0$, we have $d\Gamma_{11}=d\Gamma_{12}$ and
as a consequence that the Hugenholtz-Pines relation remains valid at any point
during the integration.

Next, we briefly turn to the regime $na\Lambda^2_{th}\ll 1$ where we put
$u_{\bf k}=1$ and $v_{\bf k}=0$ in contributions containing infrared
divergencies. In this case, only the first term in the left-hand side of Eq.\
(\ref{hupistep3}) contributes, but now with a factor 1 instead of
$u^4_\Lambda$. Thus, nothing changes with respect to the previous situation,
and the Hugenholtz-Pines relation is again valid at any point during the
renormalization if we still use the exact expressions for $u_{\bf k}$ and
$v_{\bf k}$ in contributions not containing infrared divergencies, and most
importantly in $d\tilde{n}$.

\section*{Appendix B}
\setcounter{equation}{0}
\renewcommand{\theequation}{B.\arabic{equation}}
In this appendix we discuss the implications of anomalous scaling, focussing on
the situation at the critical temperature where the coupling constants flow
into the fixed point $(\mu^*,V_{\bf 0}^*)$. We start with the fundamental
statement from renormalization group theory that the exact $n-$point vertex
function remains identical during renormalization, i.e.\ in our case
\begin{eqnarray}
\Gamma^{(n)}(p_i,\omega_i;\mu,V_{\bf 0},T;\Lambda)=
\Gamma^{(n)}(p_i,\omega_i;\mu(l),V_{\bf 0}(l),T(l);\Lambda e^{-l})~,
\end{eqnarray}
where $p_i<\Lambda e^{-l}$. The coupling constants obey the derived
renormalization group equations (without trivial scaling) and $T(l)=Te^{2l}$.
When we are at the critical temperature, we can take $\omega_i=0$ and we find
after performing the trivial rescaling that
\begin{eqnarray}
\Gamma^{(n)}(p_i,0;\mu,V_{\bf 0},T_c;\Lambda)=e^{-(3-\frac{n}{2})l}
\Gamma^{(n)}(p_i e^l,0;\mu(l),V_{\bf 0}(l),T_c(l);\Lambda)~.
\end{eqnarray}
A dimensional analysis then shows that
\begin{eqnarray}
\Gamma^{(n)}(p_i e^l,0;\mu(l),V_{\bf 0}(l),T_c(l);\Lambda)=
\Lambda^{3-\frac{n}{2}}\Gamma^{(n)}(\frac{p_i e^l}{\Lambda},0;\frac{\mu(l)}
{\Lambda^2},\frac{V_{\bf 0}(l)}{\Lambda},\frac{T_c(l)}{\Lambda^2};1)~.
\end{eqnarray}
Combining these two equations, and taking the limit $l\rightarrow\infty$ in
which we approach the fixed point, this implies
\begin{eqnarray}
\Gamma^{(n)}(p_i,0;\mu,V_{\bf 0},T_c;\Lambda)=\Lambda^{3-\frac{n}{2}}(l)
\Gamma^{(n)}(\frac{p_i}{\Lambda(l)},0;\frac{\mu^*}{\Lambda^2},\frac{V_{\bf
0}^*}{\Lambda},\infty;1)\equiv~\Lambda^{3-\frac{n}{2}}(l)
{\Gamma^{(n)}}^{*}(\frac{p_i}{\Lambda(l)})~.
\end{eqnarray}
The right-hand side has, just like the left-hand side, to be independent of
$l$, and this leads to the conclusion that
\begin{eqnarray}
{\Gamma^{(n)}}^{*}(\frac{p_i}{\Lambda(l)})=\Lambda^{\frac{n}{2}-3}(l)
{\Gamma^{(n)}}^{*}(p_i)
\end{eqnarray}
or that ${\Gamma^{(n)}}^{*}(p_i)$ has to be a homogeneous function of degree
$(3-n/2)$. Thus
\begin{eqnarray}
{\Gamma^{(n)}}^{*}(\lambda p_i)=\lambda^{3-\frac{n}{2}}{\Gamma^{(n)}}^{*}(p_i)
\end{eqnarray}
and we can conclude that anomalous scaling reveals information about the
momentum dependence of the $n-$point vertex function. In particular, we have
e.g.\ for the selfenergy at the critical temperature that
\begin{eqnarray}
{\Gamma^{(2)}}^{*}(k)\propto k^2~,
\end{eqnarray}
and most importantly for the four-point function that
\begin{eqnarray}
{\Gamma^{(4)}}^{*}({\bf k,k',K})\propto (\mid{\bf k}\mid+\mid{\bf k'}\mid)+
\alpha\mid{\bf K}\mid~,
\end{eqnarray}
which shows that the effective interaction at long wavelengths has to vanish at
the critical temperature.

Clearly, the above reasoning is in principle not restricted to the critical
temperature and a similar argument can be set up also for arbitrary
temperatures. As a result, the anomalous scaling we find from the set of
renormalization group equations derived in Sec. IVA  for the symmetry broken
phase also implies a nontrivial momentum dependence of the coupling constants.

\newpage
\section*{Figure Captions}
\noindent
Fig.\ 1 Typical behavior of the Fourier transform of the interatomic
interaction potential $V({\bf x-x'})$.\\
\\
Fig.\ 2 The one-loop Feynman graphs contributing to the renormalization of (a)
the chemical potential, (b) the two-body interaction, and (c) the three-body
interaction. The dot represents the two-body vertex $V_{\bf 0}$ and the square
the three-body vertex $U_{\bf 0}$.\\
\\
Fig.\ 3 The $p-n^{-1}-$diagram including (solid line) and excluding (dashed
line) a three-body interaction term for atomic $^{23}$Na at a temperature of
$0.1 \mu K$. The influence of $U_{\bf 0}$ is approximately 1.5 \% near the
critical density.\\
\\
Fig.\ 4 The ratio of the many-body scattering lengths $a^{RG}$ including (solid
line) and excluding (dashed line) the bubble diagrams in the flow equations as
a function of $T/T_c$ for atomic $^{23}$Na at a density of $1.5\ 10^{12}$
cm$^{-3}$.\\
\\
Fig.\ 5 Flow diagram resulting from the renormalization group equations. The
fixed point is indicated with an asterix.\\
\\
Fig.\ 6 The critical degeneracy parameter $n_c\Lambda^3_{th}$ found from the
renormalization group calculation as a function of $a/\Lambda_{th}$.\\
\\
Fig.\ 7 The one-loop Feynman diagrams contributing to the renormalization of
(a) the linear term and (b) the anomalous selfenergy in the broken phase. The
filled circle represents the vertex $V_{\bf 0}$ and the open circle the vertex
$\Gamma_3$.\\
\\
Fig.\ 8 The ratio of the effective scattering length $a^{eff}$ and the two-body
scattering length $a$ for atomic $^{23}$Na at a density of $1.5\ 10^{12}$
cm$^{-3}$. The dashed line corresponds to the result from the many-body
$T$-matrix calculation, i.e.\ $a^{eff}=a^{MB}$. The solid lines correspond to
the result from the renormalization group approach, i.e.\ $a^{eff}=a^{RG}$,
with bubble diagrams included and excluded. The effect of the bubbles (lower
solid line) is seen to be considerable, certainly below the critical
temperature.\\
\\
Fig.\ 9 The degeneracy parameter $n\Lambda^3_{th}$ as a function of the
fugacity $\zeta$ for the ideal gas (dotted line), from the many-body $T-$matrix
theory (dashed line) and as found from the renormalization group calculation
(solid line).\\
\\
Fig.\ 10 (a) The condensate fraction and (b) the superfluid fraction as a
function of temperature for a density of $1.5\ 10^{12}$ sodium atoms per cubic
centimeter, both from the many-body $T-$matrix calculation (dashed line) as
from the renormalization group calculation (solid line).\\
\\
Fig.\ 11 The pressure as a function of inverse density for atomic $^{23}$Na at
a temperature of $0.1 \mu K$. The dashed line is the result from the many-body
$T-$matrix calculation, the solid line from the renormalization group
calculation.\\
\\
Fig.\ 12 The one-loop Feynman diagrams contributing to the renormalization of
the normal selfenergy in the broken phase. The filled circle represents the
vertex $V_{\bf 0}$ and the open circle the vertex $\Gamma_3$.\\
\\

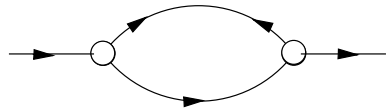
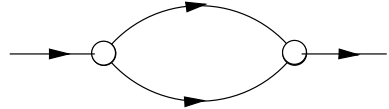
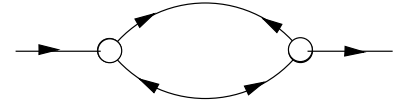
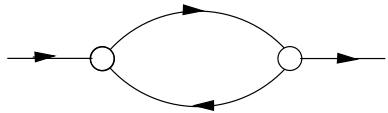
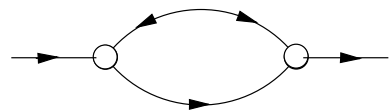
\end{document}